\documentclass[%
 aps,
 superscriptaddress,
 nofootinbib,
 balancelastpage
]{revtex4-2}

\usepackage[utf8]{inputenc}
\usepackage{amsmath,amsfonts,amssymb,graphicx,MnSymbol}
\usepackage{bbold}

\usepackage[hidelinks]{hyperref} 
\usepackage{multirow}
\usepackage{makecell}
\usepackage{tabularx}
\usepackage{siunitx}
\usepackage{physics}
\usepackage{caption}
\usepackage{subcaption}
\usepackage{cleveref} 
\usepackage[normalem]{ulem}

\crefname{section}{Sec.}{Secs.}
\crefname{table}{Table}{Tables}
\crefname{figure}{Fig.}{Figs.}
\crefname{equation}{Eq.}{Eqs.}
\crefname{appendix}{Appendix}{Appendices}

\newcommand{\ii}{\mathrm{i}} 
\newcommand{\SO}{\mathrm{SO}}
\newcommand{\SU}{\mathrm{SU}}
\newcommand{\U}{\mathrm{U}}
\newcommand{\Sp}{\mathrm{Sp}}

\def\ZZ{\mathbb{Z}_2}

\begin{document}
\raggedbottom

\title{Dark matter in composite Higgs models with a scotogenic EFT}

\author{Yu Chen}
\email{yu.chen@uni-wuerzburg.de}
\author{Werner Porod}%
 \email{werner.porod@uni-wuerzburg.de}
\affiliation{%
 Institute for Theoretical Physics and Astrophysics, Julius-Maximilians-Universit\"at W\"urzburg, 97074 W\"urzburg, Germany
}%

\begin{abstract}
We discuss a class of composite Higgs models with a fermionic UV completion that can explain the dark matter relic density. 
The resulting low-energy theory resembles so-called scotogenic models. 
A residual $\ZZ$ symmetry implies that one of the pseudo-Nambu-Goldstone bosons is stable and therefore is a viable dark matter candidate. 
As a concrete example, we take a model based on the $\SU(6)/\Sp(6)$ coset, which in principle contains four dark matter candidates.
However, only three of them can produce a relic density consistent with observations.
We perform a Markov chain Monte Carlo study focusing on dark matter observables to explore the available parameter space of the effective theory.
In particular, we find that spin-1 resonances play an important role in a substantial part of the parameter space.
Finally, we comment on possible LHC signatures of this model class.
\end{abstract}

\maketitle


\section{Introduction}

The nature of dark matter (DM) and the origin of its relic density are arguably among the most important open questions in particle physics \cite{Bertone:2018krk}.
A second open question is the origin of the fermion mass hierarchies and the intergenerational mixing in the quark and lepton sectors of the Standard Model (SM).
In particular, the lightness of neutrinos, which are several orders of magnitude lighter than the electron, the lightest electrically charged fermion, remains unexplained within the SM. The simplest solution to this puzzle is the seesaw mechanism
\cite{Minkowski:1977sc,Gell-Mann:1979vob,Yanagida:1979as,Schechter:1980gr,Cheng:1980qt,Foot:1988aq}. 
An alternative is provided by so-called scotogenic models \cite{Tao:1996vb,Ma:2006km}, which also address the DM problem.
In these models a stabilizing $\ZZ$ symmetry is introduced under which the SM particles are even. 
The additional fermions and scalars responsible for neutrino masses are odd under this symmetry. 
The lightest of the neutral $\ZZ$-odd states can serve as a DM candidate, while neutrino masses are generated at loop level only. For recent investigations of available parameter space for such models see, for example, \cite{Sarazin:2021nwo,Alvarez:2023dzz,deSouza:2025uxb} and references therein.

A third, more theoretical question is the hierarchy problem of the SM, i.e.~why is the Higgs boson so much lighter than the Planck scale. 
Composite Higgs models provide a potential solution to the hierarchy between the electroweak scale and the Planck scale by positing the Higgs boson as a composite state that originates from a new strongly interacting sector; similarly to QCD, the breaking scale is dynamically generated via confinement and condensation of a new interaction. 
This idea is as old as the SM itself \cite{Weinberg:1975gm,Dimopoulos:1979es}
starting from the first Higgsless (technicolor) theories \cite{Dimopoulos:1979za}
and their effective Lagrangian counterparts \cite{Casalbuoni:1985kq}, to models where the Higgs emerges as a composite pseudo-Nambu-Goldstone boson (pNGB) \cite{Kaplan:1983fs,Kaplan:1983sm}.
In the latter, the new strong sector has a global symmetry group $G$ which is spontaneously broken to $H$ at a scale $f \sim$ TeV. Electroweak symmetry breaking (EWSB) is induced by explicit symmetry breaking effects from electroweak (EW) gauge interactions and fermion couplings to the strong sector. 
These effects misalign the vacuum by an angle $\sin\theta = v_\mathrm{EW}/f$ \cite{Agashe:2004rs} with $v_\mathrm{EW}\simeq 246~\unit{GeV}$ corresponding to the Higgs vacuum expectation value in the SM. 
In addition, a possible mass term for the underlying fermions can also contribute to the explicit symmetry-breaking \cite{Galloway:2010bp,Cacciapaglia:2014uja}.
Realistic models with a fermionic UV completion predict a rich spectrum of composite resonances and additional pNGBs beyond the Higgs boson \cite{Ferretti:2013kya,Ferretti:2016upr,Belyaev:2016ftv}.

Strongly interacting extensions of the SM can in principle explain the observed DM relic density in various ways. In technicolor-like and composite Higgs models, there are two possibilities;
the potential DM candidate can (i) arise from the pNGB sector \cite{Frigerio:2012uc,Marzocca:2014msa,Fonseca:2015gva,Wu:2017iji,Cacciapaglia:2018avr,Alanne:2018xli,Davoli:2019tpx,Ramos:2019qqa,Cacciapaglia:2019ixa,Cai:2020njb,Rosenlyst:2021elz,Baldes:2021aph,Cornell:2022nky} or
(ii) be a technibaryon/fermionic resonance of a composite Higgs model \cite{Nussinov:1985xr,Barr:1990ca,Wu:2017iji,Cacciapaglia:2021aex,Belyakova:2024fcw,Rosenlyst:2024rsb}.
In particular, one can also obtain a scotogenic realization in which the scalars are within the pNGB sector and the $\ZZ$-symmetry is a residual symmetry of the original global symmetry $G$ \cite{Cacciapaglia:2020psm}.
We note for completeness that it could also be that the new strongly interacting sector could be completely decoupled from the SM sector or that there are only rather weak interactions between both sectors, see e.g.,~\cite{Hochberg:2014kqa,Choi:2018iit,Bernreuther:2019pfb,Appelquist:2024koa,Rosenlyst:2021elz,Bernreuther:2023kcg,Alfano:2025non,Asadi:2026mip}, leading to dark sector/hidden valley models.

In the present paper we focus on dark matter aspects of composite Higgs models which have a residual $\ZZ$ symmetry in the effective theory as needed for scotogenic models. 
As a concrete example we take the model presented in Ref.~\cite{Cacciapaglia:2020psm}. 
We have identified four possible DM candidates among the $\ZZ$-odd pNGBs of which, however, only three can explain the observed relic density.
We explore the available parameter space by performing Markov chain Monte Carlo (MCMC) scans. 
In this context, we find that interactions of the $\ZZ$-odd pNGBs with the Higgs boson as well as quartic interactions of $\ZZ$-odd pNGBs with two electroweak vector bosons are important. 
Moreover, it turns out that also heavy spin-1 resonances are relevant in the calculation of the dark matter relic density. 
A large portion of the parameter space, in which the observed dark matter relic density can be explained, is excluded by direct dark matter searches.  

The paper is organized as follows. We present the relevant details of the model considered in Sec.~\ref{sec:model}. In Sec.~\ref{sec:dm_observables} we discuss first the relevant contributions to the various (co)annihilation processes before discussing some specific features of the calculation. 
Afterwards, we perform a MCMC study to explore the allowed region in parameter space for all possible DM candidates.  Moreover, we present results on the constraints from dark matter direct detection. 
Subsequently, in Sec.~\ref{sec:lhc} we summarize how this class of models can be tested at the LHC or a future FCC-hh. Finally, we draw our conclusions in Sec.~\ref{sec:onclusions}. In addition, we collect some technical details of the model in the Appendix.

\section{Model aspects}
\label{sec:model}

In this section, we summarize the main aspects of the model presented in \cite{Cacciapaglia:2020psm} relevant to this work. In addition, we will present some details on the scalar potential and on the spin-1 resonances which will be needed later. 

The model postulates the existence of six hyper-fermions at very high energies which are charged under an $\Sp(4)$ gauge group $G_{HC}$ in addition to the electroweak SM gauge group.
It is assumed that the $\Sp(4)$ gauge interactions become strong in the multi-TeV region, leading to the formation of bound states. The hyperfermions form a fundamental $\mathbf{6}$ of an $\SU(6)$ global group if one neglects the electroweak gauge interactions. At the condensation scale, the $\SU(6)$ gets spontaneously broken to $\Sp(6)$ by the condensate. The unbroken $\Sp(6)$ contains the custodial group $ \SU(2)_L\times \SU(2)_R$, as well as a $\ZZ$ symmetry, which commute with each other, see~\cite{Cacciapaglia:2020psm} for further details. One identifies the hypercharge with the third generator of SU(2)$_R$. 
The gauging of SU(2)$_L\times$U(1)$_Y$ explicitly breaks the global SU(6) symmetry and eventually contributes to the scalar potential. We give the electroweak and $\ZZ$ quantum numbers of the hyperfermions in Table \ref{tab:embedding}. In terms of the sixplet of Weyl spinors, $\Psi$, the underlying fermionic Lagrangian reads as
\begin{align}
 {\cal L}_{\Psi} = \bar{\Psi} \ii \gamma^\mu D_\mu \Psi - \frac{1}{2}  \left( \Psi^T M_\Psi \Psi + \mathrm{H.c.} \right) + \delta {\cal L}
\end{align}
where the covariant derivatives include the $G_{HC}$, $\SU(2)_L$, and $\U(1)_Y$ gauge
bosons. The mass term consists of three independent masses
\begin{align}
  M_\Psi = \mathrm{Diag}\left(\ii m_1 \sigma_2, -\ii m_2 \sigma_2,\ii m_3 \sigma_2 \right)  \,.
\label{eq:baremass}
\end{align}
In case of $m_1=m_2=m_3$ one has an explicit breaking of the global $\SU(6)$ to $\Sp(6)$ whereas if the masses are different, the $\SU(6)$ breaks to $\SU(2)_L\times \SU(2)_R$ \cite{Cacciapaglia:2020psm}.
The additional terms in $\delta {\cal L}$ contain higher-dimensional operators responsible for
generating masses for the SM fermions in the condensed phase as well as for parts of the scalar potential.

\begin{table}[t]
\captionsetup{justification=raggedright}
\begin{center}
    \begin{tabular}{ |c|c|c|}
     \hline
     Hyperfermions & $\SU(2)_{L} \times \SU(2)_{R}$ &  $\ZZ$\\
    \hline
     $\Psi_1 \equiv (\psi_1,\hspace{0.1cm}\psi_2)^T$ & (\textbf{2},\textbf{1}) & $+$ \\ 
     $\Psi_2 \equiv (\psi_3,\hspace{0.1cm}\psi_4)^T$ & (\textbf{1},\textbf{2}) & $+$ \\ 
     $\Psi_3 \equiv (\psi_5,\hspace{0.1cm}\psi_6)^T$ & (\textbf{2},\textbf{1})  & $-$\\ 
     \hline
    \end{tabular}
 \end{center}
  \caption{\label{tab:embedding} Hyperfermions and their SU(2) and $\ZZ$ quantum numbers.
    }
\end{table}  

\subsection{The pNGB sector}

The spontaneous breaking of $\SU(6)$ to $\Sp(6)$ gives rise to 14 pNGBs among which are the four degrees of freedom needed for the Higgs doublet of the SM. They belong to the $\ZZ$-even sector which in addition contains two electroweak gauge singlets, $\eta_1$ and $\eta_2$. The latter are pseudoscalars that have anomaly-induced couplings to the electroweak gauge bosons. 
The other additional pNGBs including the $\ZZ$-odd ones do not have any anomaly-induced couplings to two vector bosons.
The $\ZZ$-odd sector consists of an $\SU(2)_L$ triplet $\Delta$, an $\SU(2)_L\times\SU(2)_R$ bidoublet $\Phi$, and a gauge singlet $\eta_3$. The quantum numbers of all pNGB states are summarized in Table \ref{tab:spectrum}. The unbroken $\ZZ$ symmetry makes the lightest $\ZZ$-odd state stable. Since the $\ZZ$-odd sector contains four neutral fields, $\Delta^0$, $\phi^0$, $\eta_3$, and $\eta_4$, each can in principle become the lightest stable state and thus serve as a DM candidate. 

\begin{table}[th]
\captionsetup{justification=raggedright}
\begin{center}
    \renewcommand{\arraystretch}{1.2} 
    \begin{tabular}{ |c|c c c|c c c|c c c| } 
    \hline
    & \multicolumn{3}{c|}{pNGBs} & \multicolumn{3}{c|}{$\mathcal{A}_\mu$} & \multicolumn{3}{c|}{$\mathcal{V}_\mu$} \\
    & name & $\text{SU(2)}^2$ &  $\text{SU(2)}_{\rm V}$
   & name & $\text{SU(2)}^2$ &  $\text{SU(2)}_{\rm V}$
     & name & $\text{SU(2)}^2$ &  $\text{SU(2)}_{\rm V}$\\
    \hline
    \multirow{5}{*}{$\mathbb{Z}_2$-even}  
        & $\varphi$ & \multirow{2}{*}{(2,2)} & 3 
        & $a_{1\mu}$ & \multirow{2}{*}{(2,2)} & 3 &
        $v_{1\mu}$ & $(3,1)\oplus(1,3)$ & 3 \\
        & $h$ & & 1 & $y_{1\mu}$ & & 1 & $v_{2\mu}$ & $(3,1)\oplus(1,3)$ & 3 \\
        & $\eta_1$ &(1,1) & 1  & $y_{2\mu}$ &(1,1) & 1 & $v_{3\mu}$ & (3,1) & 3 \\
         & $\eta_2$ &(1,1) & 1 & $y_{3\mu}$ & (1,1) & 1& $r_{1\mu}$ & \multirow{2}{*}{(2,2)}& 3  \\
         & & & & & & & $x_{1\mu}$ & & 1\\               
    \hline
    \multirow{4}{*}{$\mathbb{Z}_2$-odd}  & $\Delta$ & (3,1) & 3 & $a_{2\mu}$ & (3,1) & 3&   
        $r_{2\mu}$ & (3,1) & 3 \\
       & $\eta_3$ & (1,1) & 1 & $y_{4\mu}$ & (1,1) & 1 & $x_{2\mu}$ & (1,1) & 1 \\
        & $\phi$ &\multirow{2}{*}{(2,2)} & 3  & $a_{3\mu}$ &\multirow{2}{*}{(2,2)} & 3  & $r_{3\mu}$ & \multirow{2}{*}{(2,2)} & 3\\
      & $\eta_4$ & & 1  & $y_{5\mu}$ & & 1  & $x_{3\mu}$ & & 1 \\      
    \hline
\end{tabular}
\end{center}
\caption{\label{tab:spectrum} pNGBs and spin-1 resonances and their quantum numbers; the $\varphi$ are the would-be Goldstone bosons forming the longitudinal components of the electroweak vector bosons.}
\end{table}

The self-interactions of the pNGBs have two sources; (i) interactions with two space-time derivatives from the Callan-Coleman-Wess-Zumino (CCWZ) construction \cite{Coleman:1969sm,Callan:1969sn}, and (ii) interactions from the effective potential. 
The quartic scalar interactions from the CCWZ construction scale like $p_1 p_2/f^2_\pi$ with $p_i$
being the four momenta of two scalars and $f_\pi$ the
decay constant. 
These terms are in general subleading for the relic density calculation but become important
if the corresponding interactions from the scalar potential are small.
The effective potential of the composite scalars receives contributions from EW gauge interactions, SM fermion couplings to the strong sector, and the hyperfermion masses. 
Reference~\cite{Alanne:2018wtp} provides a general classification of the template operators, up to next-to-leading order, that appear in chiral perturbation theories based on the spontaneous symmetry breaking patterns $\SU(N)/\Sp(N)$ and $\SU(N)/\SO(N)$. Already in the simplest case of $\SU(4)/\Sp(4)$, it has been found that there are 11 independent low-energy parameters characterizing the effective potential assuming $CP$ conservation \cite{Golterman:2017vdj}. In the model at hand, many more free parameters will be needed to characterize the effective potential.

In this paper we will investigate four potential DM candidates $\Delta^0$, $\phi^0$, $\eta_3$, and $\eta_4$. 
We will make some assumptions to reduce the number of free parameters. 
(i) We treat the pNGB masses as free parameters. The exception is of course the mass of the Higgs boson $h$ which we fix to 125~GeV. 
We will further assume that most pNGBs are sufficiently heavier than the DM candidate that they are not relevant in the relic density calculation;
(ii) The only spin-0 particle related to electroweak symmetry breaking is the Higgs boson;
(iii) We do not consider any mixing between the various states. 
Such mixing would change details of the DM calculations but not the overall features;
(iv) We reduce the number of free parameters in the quartic part of the potential assuming one were to start from an $\SU(2)_L \times \U(1)_Y$ invariant theory before electroweak symmetry breaking, and denote the multiplet from which $\phi$ and $\eta_4$ would emerge by $\Phi$. 
Under this assumption, the relevant part of the scalar potential reads as
\begin{align}
    \mathcal{L}_\mathrm{pot} \supset &~
    \hat{c}_1\eta_3^2 \Tr(\Delta\Delta) +     
    \hat{c}_2 \Tr(\Delta\Delta\Delta\Delta) + 
    \hat{c}_3 \eta_3^2 \Tr(H^\dagger H) \nonumber + 
    \hat{c}_4 \Tr(\Delta\Delta)\Tr(H^\dagger H) \\
    & + \hat{c}_5 \Tr(H^\dagger\Phi)\Tr(\Phi^\dagger H) +
    \hat{c}_6 \Tr(\Phi^\dagger\Phi)\Tr(H^\dagger H) + 
    \hat{c}_7 \big[\Tr(\Phi^\dagger\Phi)\big]^2.
\end{align}
with $H$ being the multiplet that contains the SM Higgs doublet.
We note for completeness that $\Tr(\Delta\Delta\Delta\Delta)=  \Tr(\Delta\Delta)^2/2$ and that $\Tr(\Delta\Delta)\Tr(H^\dagger H) = \Tr(H^\dagger \Delta\Delta H)/2$ in the unitary gauge.
In terms of the actual states, the relevant part of the potential is given by 
\begin{align}
    \mathcal{L}_\mathrm{pot} \supset &~
    c_1 \eta_3^2 \Big[(\Delta^0)^2 + 2\Delta^+\Delta^- \Big] +    
    c_2 \Big[(\Delta^0)^4 + 4(\Delta^0)^2 \Delta^+\Delta^- + 4(\Delta^+\Delta^-)^2 \Big] \nonumber\\
    & + c_3 \eta_3^2 \big(h^2 + 2 v_{\rm EW} h\big) + 
    c_4 \Big[(\Delta^0)^2 + 2\Delta^+\Delta^- \Big]\big(h^2 + 2 v_{\rm EW} h\big) \nonumber\\
    &+ c_5 \eta_4^2 \big(h^2 + 2 v_{\rm EW} h\big) + c_6 \Big[(\phi^0)^2 + 2\phi^+\phi^- \Big]\big(h^2 + 2 v_{\rm EW} h\big) \nonumber\\
    &+ c_7 \Big[\eta_4^4 + 2\eta_4^2 (\phi^0)^2 + 4\eta_4^2 \phi^+\phi^- + (\phi^0)^4 + 4(\phi^0)^2 \phi^+\phi^- + 4(\phi^+\phi^-)^2 \Big] .
\label{eq:Veff}    
\end{align}
The coefficients $c_i$ are free parameters of the model. Clearly this a simplifying assumption, and we expect that within a given square bracket there will be additional relative coefficients that are close to one, e.g.,~of the size of $\cos\theta$, in front of different terms.
Furthermore, we assume $c_2,\;c_7>0$ to ensure that the scalar potential is bounded from below\footnote{This is a necessary but not sufficient condition. A complete analysis of vacuum stability, including the possible existence of charge-breaking minima, would require to consider the full scalar potential and is beyond the scope of this work.}.
For simplicity, we set the Higgs trilinear and quartic self-couplings to their SM values.
This leaves us with the pNGB masses as additional free parameters.
Also here we make some simplifying assumptions for the pNGB masses: (i) The DM candidate is the lightest of the additional pNGBs beyond the Higgs boson and we denote its mass by $M_\chi$ independent of its nature; 
(ii) For the next heavier $\ZZ$-odd pNGBs that participate in coannihilation, we assign a common mass $M_\chi + \Delta M$;
(iii) All other pNGBs are assumed to be so heavy that they do not play a role here.

\subsection{Spin-1 resonances}
\label{subsec:spin_one}

Composite Higgs models also predict the existence of composite spin-1 resonances. 
Some of them can play an important role in the relic density calculation, as will be discussed below. They come in two varieties: vectors and axial-vectors.
The main difference between these two types is that vectors can couple to two pNGBs and axial-vectors only couple to three pNGBs before any mixing with the electroweak vector bosons is taken into account. 
We summarize them and their quantum numbers also in Table \ref{tab:spectrum}.

In the absence of a bare mass for the hyperfermions, i.e.,~for $m_1=m_2=m_3=0$, one would expect that the spin-1 resonances are significantly heavier than the pNGBs, implying that they could be neglected in the DM calculation. We do not want to make this assumption on the masses of the hyperfermions. In such a case one still expects that the spin-1 resonances are heavier than the pNGBs but with mass ratios of as low as two,
as indicated by both lattice studies \cite{Bennett:2019cxd,Kulkarni:2022bvh,Bennett:2023wjw,Bennett:2024tex}  and gauge/gravity duality calculations \cite{Erdmenger:2020flu,Erdmenger:2024dxf,Alfano:2025non}. We will thus take the overall mass scale $M_V$ of the spin-1 resonances as a free parameter. More precisely, this is the mass parameter of the vector states related to the unbroken $\Sp(6)$ generators.
The corresponding mass parameter of the axial-vectors, the states related to the broken $\SU(6)$ generators, is denoted by $M_A$.
We have used the hidden symmetry approach \cite{Bando:1987br} to calculate the mass matrices and the interactions of the spin-1 resonances
following refs.~\cite{BuarqueFranzosi:2016ooy,Caliri:2024jdk}.
We summarize here the main aspects relevant for DM calculations and refer to Appendix~\ref{app:hidden_symmetry} for the technical details.

The electroweak gauge bosons mix with some of the $\ZZ$-even resonances. In the charged sector, the electroweak eigenstate $\tilde W^\pm$ mixes with the axial-vector state $a_1^\pm$ and with a linear combination of the vector states $v_1^\pm$, $v_2^\pm$, and $v_3^\pm$, denoted in the following by $V_1^\pm$. The resulting mass eigenstates are
$R_\mu^\pm =
(W_\mu^\pm,\allowbreak
A_{1,\mu}^\pm,\allowbreak
V_{1,\mu}^\pm,\allowbreak
V_{2,\mu}^\pm,\allowbreak
V_{3,\mu}^\pm)^T$ where the mixing leads to physical masses different from the underlying mass parameters $M_A$ and $M_V$.
In the neutral sector, the electroweak states $B_\mu$ and $\tilde W^3_\mu$ mix with $a_1^0$ and with two linear combinations of $v_1^0$, $v_2^0$, and $v_3^0$, denoted by $V_1^0$ and $V_2^0$. This gives the mass eigenstates
$R_\mu^0 =
(a_\mu,\allowbreak Z_\mu,\allowbreak
A_{1,\mu}^0,\allowbreak
V_{1,\mu}^0,\allowbreak
V_{2,\mu}^0,\allowbreak
V_{3,\mu}^0)^T$.
The corresponding mass matrices are given in Appendix~\ref{app:hidden_symmetry}.
The states $V_1^{\pm,0}$ are essentially a $\SU(2)_L$ multiplet mixing sizeably with the SM $\SU(2)_L$ gauge bosons and $V_2^0$ is the neutral component of a $\SU(2)_R$ triplet which mixes sizeably with the SM $\U(1)_Y$ boson. This is very similar to the case of the minimal cosets discussed in refs.~\cite{BuarqueFranzosi:2016ooy,Caliri:2024jdk}.
This also implies that the limits on mass-coupling combinations due to existing collider data will be very similar to those presented in \cite{Caliri:2024jdk}.

This formalism gives also the couplings between the pNGBs and the spin-1 resonances. The couplings depend on a few model parameters, as discussed below, and the mixing matrices in the spin-1 sector. The corresponding couplings are lengthy expressions. 
Thus, we indicate here the important ones and sort them according to their Lorentz structures:
\begin{itemize}
 \item $h$-$\chi$-$V_-$, $h$-$\pi$-$V_-$, $\chi$-$\pi$-$V_+$
 \item $\chi$-$V^0_-$-$R^0$, $\chi$-$V^\pm_-$-$R^\mp$, $\pi$-$V_-$-$R^0$, $\pi$-$V_-$-$R^\mp$, and $h$-$W^+$-$W^-$, $h$-$Z$-$Z$, $h$-$Z$-$R^0$, $h$-$W^\pm$-$R^\mp$
 \item $\chi$-$A_-^0$-$R^0$, $\chi$-$A_-^\mp$-$R^\pm$, $\pi$-$A_-$-$R^0$, $\pi$-$A_-$-$R^\pm$ and $h$-$A^0_+$-$R^0$, $h$-$A^\mp_+$-$R^\pm$
 \item $\chi$-$\chi$-$R^+$-$R^-$, $\chi$-$\chi$-$R^0$-$R^0$, $\chi$-$\pi$-$V$-$V'$, $\pi$-$\pi$-$V$-$V'$ ($V,V' = R^\pm, R^0$)
\end{itemize}
Here, $\chi$ denotes the DM candidate, while $\pi$ represents any of the heavier $\ZZ$-odd pNGBs. Furthermore, $V_-/A_-$ ($V_+/A_+$) denotes any of the $\ZZ$-odd ($\ZZ$-even) spin-1 states. These interactions contribute to both direct DM annihilation and coannihilation processes. For the contact interactions shown in the last line, we retain only vertices involving at least one electroweak gauge boson, since annihilation into a pair of heavy vector resonances is kinematically inaccessible in the relic density calculation.

Moreover, the mixing of the spin-1 resonances with the SM gauge bosons implies couplings of these states with the SM fermions. 
The corresponding interaction Lagrangian is given by
\begin{align}
    \mathcal{L}_{\rm CC} & = \frac{\hat{g}}{\sqrt{2}} \sum_{i,f}\mathcal{C}_{1i}\Bar{\psi}_f R_\mu^+\gamma^\mu P_L\psi_f + \rm H.c.,\\
    \mathcal{L}_{\rm NC} & = \sum_{i,f}\Bar{\psi}_f R_\mu^0\gamma^\mu (g_{Li}^f P_L + g_{Ri}^f P_R)\psi_f,
\end{align}
with\begin{equation}
    g_{Li}^f = \hat{g} T_f^3 \mathcal{N}_{2i} + \hat{g}^\prime Y_{f_R} \mathcal{N}_{1i} \quad\text{and}\quad g_{Ri}^f = \hat{g}^\prime Y_{f_R} \mathcal{N}_{1i}.
\end{equation}
Here, $T_f^3$ is the weak isospin charge of the fermion $f$, and $Y_{f_L/f_R}$ is the corresponding hypercharges. And $\mathcal{C}$ and $\mathcal{N}$ denote the mixing matrices in the charged and neutral sectors, respectively. Details can be found in Appendix \ref{app:hidden_symmetry}.
We note for completeness that in case of third generation quarks one gets in principle additional contributions from the partial compositeness approach to explain the large top mass. 
However, we neglect them here for simplicity.

\subsection{Independent parameters}

We will investigate the four possible DM candidates in the following. 
It is well known that coannihilation channels can be quite important in addition to direct DM annihilation in determining the relic density \cite{Griest:1990kh}.
We will consider the following two combinations of pNGBs to account for this effect while keeping the analysis simple:
(i) $\eta_3$ and $\Delta$ and (ii) $\eta_4$ and $\phi$. 
In both cases, the DM mass $M_\chi$ and the mass splitting $\Delta_M$ between the DM candidate and the three next heavier coannihilating states are treated as free parameters. 
For the first (second) combination, the effective potential in Eq.~\ref{eq:Veff} contains four (three) independent free couplings. 
We denote the quartic coupling between the DM candidate and the Higgs boson by $c_{\chi\chi hh}$, while the corresponding coupling involving the coannihilating pNGBs is denoted by $c_{\pi\pi hh}$. 
They are alternatively $c_3$ or $c_4$ in Eq.~\ref{eq:Veff} for combination (i) and $c_5$ or $c_6$ for combination (ii).

In addition, there are eight parameters in the spin-1 sector and its interactions with the pNGBs, see Appendix~\ref{app:hidden_symmetry}. 
Three of them are fixed by the fine structure constant $\alpha$, the mass of the $Z$-boson and the Fermi constant $G_F$. 
This leaves us with five independent parameters in this sector which we have chosen as follows. 
The coupling $\tilde g$ between the spin-1 resonances, 
the mass parameter $M_V$ for the heavy vector states, $\xi=M_A/M_V$ with $M_A$ being the mass parameter for the axial vectors, 
the coupling $g_{V\pi\pi}$ of two pNGBs with a heavy spin-1 vector in the limit of vanishing electroweak gauge couplings, 
and the misalignment angle $\theta$. 

\section{DM observables}
\label{sec:dm_observables}

The model under consideration contains four potential cold DM candidates in the pNGB sector, $\eta_3$, $\Delta^0$, $\eta_4$, and $\phi^0$.
As mentioned above, we consider simplified scenarios with two combinations of pNGB states: (i) $\eta_3$ and $\Delta$; (ii) $\eta_4$ and $\phi$.
In a weakly coupled model, the first combination would correspond to a gauge singlet together with a real $\SU(2)_L$ triplet, while the second would correspond to an inert Higgs doublet.

\begin{figure}[ht]
\captionsetup{justification=raggedright}
    \centering     
    \includegraphics[width=0.7\textwidth]{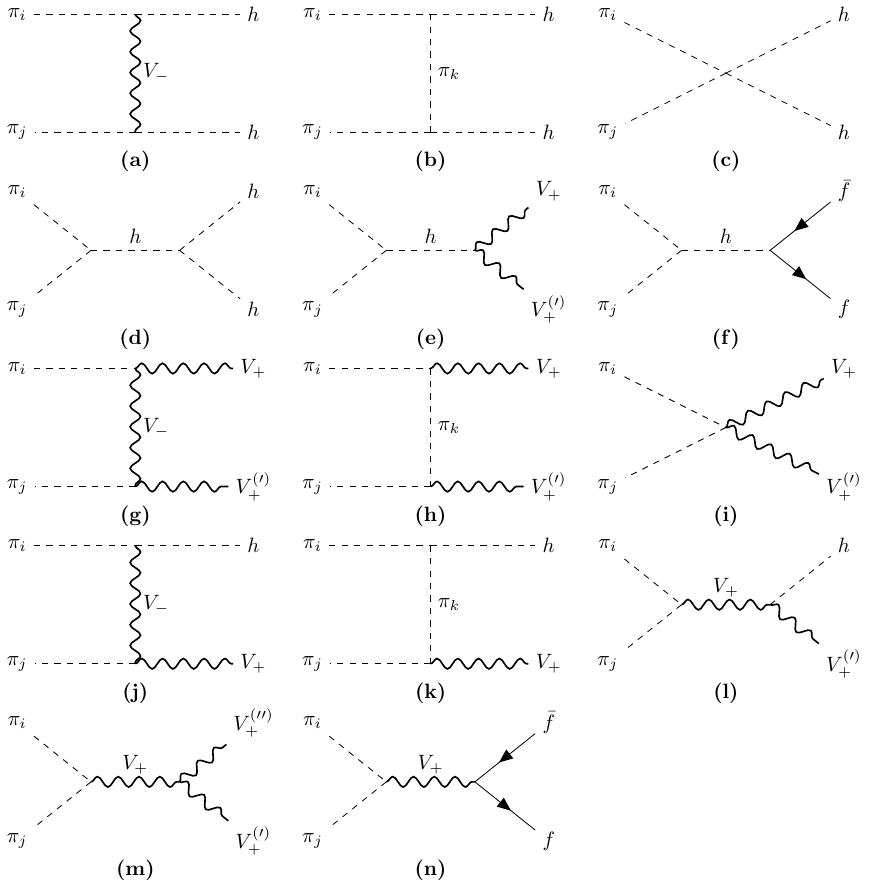}
    \caption{Topologies of (co)annihilation processes relevant for the DM relic density calculation in the model. Here, $\pi_{i,j,k}$ denote $\mathbb{Z}_2$-odd pNGBs. The final-state vector bosons correspond to the $\mathbb{Z}_2$-even states contained in $R_\mu^\pm$ and $R_\mu^0$, while $f\bar f$ includes all SM fermion pairs. Topologies (a)–(i) contribute to both annihilation and coannihilation processes, whereas topologies (j)–(n) arise exclusively in coannihilation channels, due to the absence of vertices of the form $\pi \partial_\mu \pi V^\mu$ for two identical pNGBs.}
    \label{fig:anni}
    \end{figure}

The DM phenomenology is controlled by interactions among the $\ZZ$-odd pNGBs, and their interactions with the SM fields and with spin-1 resonances discussed in the previous section.
These interactions contribute both to the annihilation and coannihilation processes that determine the relic abundance and to the scattering processes relevant for direct detection experiments. The corresponding topologies of annihilation and coannihilation are shown in Fig.~\ref{fig:anni}, among which some also contribute to the scattering. The relevant mediators for different processes of the two combinations under consideration are summarised in Table~\ref{tab:mediatorsEtaDelta} for $\eta_3$ and $\Delta$, and Table~\ref{tab:mediatorsPhi} for $\eta_4$ and $\phi^0$.

\begin{table}[t]
\captionsetup{justification=raggedright}
    \centering \scriptsize 
    \setlength{\tabcolsep}{2.5pt} 
    \renewcommand{\arraystretch}{1.15} 
    \resizebox{\textwidth}{!}{%
  \begin{tabular}{|c|c|c|c|c|c|c|c|} 
  \hline
   \multirow{2}{*}{final state} & \multicolumn{7}{c|}{initial state} \\
   \cline{2-8}
     &  $\eta_3 \eta_3$ &  $\eta_3 \Delta^0$ & $\eta_3 \Delta^\pm$ &  $\Delta^0\Delta^0$ & $\Delta^0\Delta^\pm$ & $\Delta^+\Delta^-$& $\Delta^\pm\Delta^\pm$ \\ \hline 
   $hh$ & \makecell[l]{(a) $x_3$\\ (b) $\eta_3$\\ (c), (d)} & ---  & --- & \makecell[l]{(a) $r_3^0$\\ (b) $\Delta^0$\\ (c), (d)} & --- & \makecell[l]{(a) $r_3^\pm$\\ (b) $\Delta^\pm$\\ (c), (d)} & ---\\ \hline
   $WW$ & \makecell[l]{(g) $a_2^\pm$, $r_3^\pm$\\ (h) $\Delta^\pm$\\ (e), i)} & \makecell[l]{(g) $a_2^\pm$, $r_3^\pm$\\ (h) $\Delta^\pm$\\ (m) $R^0$} & --- & \makecell[l]{(g) $a_2^\pm$, $r_3^\pm$\\ (h) $\Delta^\pm$\\ (e), (i)} & --- & \makecell[l]{(g) $a_2^0$, $r_3^0$, $y_4$, $x_3$\\ (h) $\Delta^0$, $\eta_3$\\ (m) $R^0$ \\ (e), (i)} & \makecell[l]{(g) $a_2^0$, $r_3^0$, $y_4$, $x_3$\\(h) $\Delta^0$, $\eta_3$\\ (i)} \\ \hline
   $ZZ$ & \makecell[l]{(g) $a_2^0$, $r_3^0$\\ (h) $\Delta^0$\\ (e), (i)} & --- & --- & \makecell[l]{(g) $y_4$, $x_3$\\ (h) $\eta_3$\\ (e), (i)} & --- & \makecell[l]{(g) $a_2^\pm$, $r_3^\pm$\\ (h) $\Delta^\pm$\\ (e), (i)}  & --- \\\hline
   $\bar{f} f^{(')}$ & (f) & (n) $R^0$ & (n) $R^\pm$ & (f) &  (n) $R^\pm$ & (n) $R^0$ & --- \\ \hline
   $hZ$ & --- & \makecell[l]{(j) $r_3^0$ ,$x_3$\\ (k) $\Delta^0$, $\eta_3$\\ (l)$R^0$} & --- & --- & --- & \makecell[l]{(j) $r_3^\pm$\\ k) $\Delta^\pm$\\ (l) $R^0$} & ---  \\\hline
   $h\gamma$ & --- & --- & --- & --- & --- & (k) $\Delta^\pm$ & ---  \\\hline
   $hW$ & --- & --- & \makecell[l]{(j) $r_3^\pm$, $x_3$\\ (k) $\Delta^\pm$, $\eta_3$\\ (l) $R^\pm$} & --- & \makecell[l]{(j) $r_3^{\pm,0}$\\ (k) $\Delta^{\pm,0}$\\ (l) $R^\pm$} & --- & ---  \\\hline
   $ZW$ & --- & --- & \makecell[l]{(g) $a_2^{\pm,0}$, $r_3^{\pm,0}$\\ (h) $\Delta^{\pm,0}$\\ (m) $R^\pm$ \\ (i)} & --- & \makecell[l]{(g) $a_2^\pm$, $r_3^\pm$, $y_4$, $x_3$\\ (h) $\Delta^\pm$, $\eta_3$\\ (m) $R^\pm$ \\ (i)} & --- & ---  \\\hline
   $\gamma W$ & --- & --- & \makecell[l]{(h) $\Delta^{\pm}$\\ (m) $R^\pm$ \\ (i)} & --- &  \makecell[l]{(h) $\Delta^{\pm}$\\ (m) $R^\pm$ \\ (i)} & --- & ---  \\\hline
    \end{tabular}}
    \caption{Intermediate states contributing to different processes in Fig.~\ref{fig:anni} sorted according to the initial states and final states in the $\eta_3/\Delta$ scenario. $R^0$ and $R^\pm$ are the electroweak vector bosons and the spin-1 resonances mixing with them, see Sec. \ref{subsec:spin_one}. Furthermore, $f\bar f^{(\prime)}$ includes all SM fermion pairs.}
    \label{tab:mediatorsEtaDelta}
\end{table}

\begin{table}[t]
\captionsetup{justification=raggedright}
    \centering \scriptsize 
    \setlength{\tabcolsep}{2.5pt} 
    \renewcommand{\arraystretch}{1.15} 
    \resizebox{\textwidth}{!}{%
  \begin{tabular}{|c|c|c|c|c|c|c|c|} 
  \hline
   \multirow{2}{*}{final state} & \multicolumn{7}{c|}{initial state} \\
   \cline{2-8}
     &  $\eta_4 \eta_4$ &  $\eta_4 \phi^0$ & $\eta_4 \phi^\pm$ &  $\phi^0\phi^0$ & $\phi^0\phi^\pm$ & $\phi^+\phi^-$& $\phi^\pm\phi^\pm$ \\ \hline 
   $hh$ & \makecell[l]{(a) $x_2$\\ (b) $\eta_4$\\ (c), (d)} & ---  & --- & \makecell[l]{(a) $r_2^0$\\ (b) $\phi^0$\\ (c), (d)} & --- & \makecell[l]{(a) $r_2^\pm$\\ (b) $\phi^\pm$\\ (c), (d)} & ---\\ \hline
   $WW$ & \makecell[l]{(g) $a_3^\pm$, $r_2^\pm$\\ (h) $\phi^\pm$\\ (e), (i)} & \makecell[l]{(g) $a_3^\pm$, $r_2^\pm$\\ (h) $\phi^\pm$\\ (m) $R^0$} & --- & \makecell[l]{(g) $a_3^\pm$, $r_2^\pm$\\ (h) $\phi^\pm$\\ (e), (i)} & --- & \makecell[l]{(g) $a_3^0$, $r_2^0$, $y_5$, $x_2$\\ (h) $\phi^0$, $\eta_4$\\ m) $R^0$ \\ (e), (i)} & \makecell[l]{(g) $a_3^0$, $r_2^0$, $y_5$, $x_2$\\ (h) $\phi^0$, $\eta_4$\\ (i)} \\ \hline
   $ZZ$ & \makecell[l]{(g) $a_3^0$, $r_2^0$\\ (h) $\phi^0$\\ (e), (i)} & --- & --- & \makecell[l]{(g) $y_5$, $x_2$\\ (h) $\eta_4$\\ (e), (i)} & --- & \makecell[l]{(g) $a_3^\pm$, $r_2^\pm$\\ (h) $\phi^\pm$\\ (e), (i)}  & --- \\\hline
   $\bar{f} f^{(')}$ & (f) & (n) $R^0$ & (n) $R^\pm$ & (f) &  (n) $R^\pm$ & (n) $R^0$ & --- \\ \hline
   $hZ$ & --- & \makecell[l]{(j) $r_2^0$ ,$x_2$\\ (k) $\phi^0$, $\eta_4$\\ (l) $R^0$} & --- & --- & --- & \makecell[l]{(j) $r_2^\pm$\\ (k) $\phi^\pm$\\ (l) $R^0$} & ---  \\\hline
   $h\gamma$ & --- & --- & --- & --- & --- & (k) $\phi^\pm$ & ---  \\\hline
   $hW$ & --- & --- & \makecell[l]{(j) $r_2^\pm$, $x_2$\\ (k) $\phi^\pm$, $\eta_4$\\ (l) $R^\pm$} & --- & \makecell[l]{(j) $r_2^{\pm,0}$\\ (k) $\phi^{\pm,0}$\\ (l) $R^\pm$} & --- & ---  \\\hline
   $ZW$ & --- & --- & \makecell[l]{(g) $a_3^{\pm,0}$, $r_2^{\pm,0}$\\ (h) $\phi^{\pm,0}$\\ (m) $R^\pm$ \\ (i)} & --- & \makecell[l]{(g) $a_3^\pm$, $r_2^\pm$, $y_5$, $x_2$\\ (h) $\phi^\pm$, $\eta_4$\\ (m) $R^\pm$ \\ (i)} & --- & ---  \\\hline
   $\gamma W$ & --- & --- & \makecell[l]{(h) $\phi^{\pm}$\\ (m) $R^\pm$ \\ (i)} & --- &  \makecell[l]{(h) $\phi^{\pm}$\\ (m) $R^\pm$ \\ (i)} & --- & ---  \\\hline
    \end{tabular}}
    \caption{Intermediate states contributing to different processes in Fig. \ref{fig:anni} sorted according to the initial states and final states  in the $\eta_4/\phi$ scenario. $R^0$ and $R^\pm$ are the electroweak SM vector bosons and the spin-1 resonances mixing with them, see Sec. \ref{subsec:spin_one}. Furthermore, $f\bar f^{(\prime)}$ includes all SM fermion pairs. }
    \label{tab:mediatorsPhi}
\end{table}

\subsection{Relic density}\label{sec:RD}

The present abundance of cold DM is determined by thermal freeze-out in the early Universe and therefore depends on the annihilation and coannihilation cross sections of the $\ZZ$-odd pNGB states. 
To evaluate these processes, we implemented the model in \texttt{FeynRules} \cite{Alloul:2013bka} and generated the corresponding \texttt{CalcHEP} \cite{Belyaev:2012qa} model files. These files were then interfaced with \texttt{micrOMEGAs~6.2} \cite{Belanger:2013oya,Alguero:2023zol}\footnote{The corresponding codes are available from the authors upon request.}, which was used to compute the DM relic density and the direct detection cross sections of the model for comparison with experimental measurements.

The relic density has been determined rather precisely by the Planck Collaboration~\cite{Planck:2018vyg}:
\begin{equation}
    \Omega_c h^2 = 0.120 \pm 0.001\,.
   \label{eq:dm_range_exp} 
\end{equation}
In the following, we do not require the relic density obtained for an allowed parameter point to lie within the experimental $1\sigma$ range, but instead impose $\Omega_c h^2 = 0.120 \pm 0.012$ as a bound. 
There are three reasons for this: 
(i) We make some simplifying assumptions to reduce the number of free parameters in the scalar potential. 
(ii) We do not take into account potential contributions from other resonances in the calculations of the cross sections. 
(iii) We use only lowest order calculations.

\begin{figure}[t]
\captionsetup{justification=raggedright}
    \centering
    \begin{subfigure}{0.48\linewidth}
        \centering
        \includegraphics[width=\linewidth]{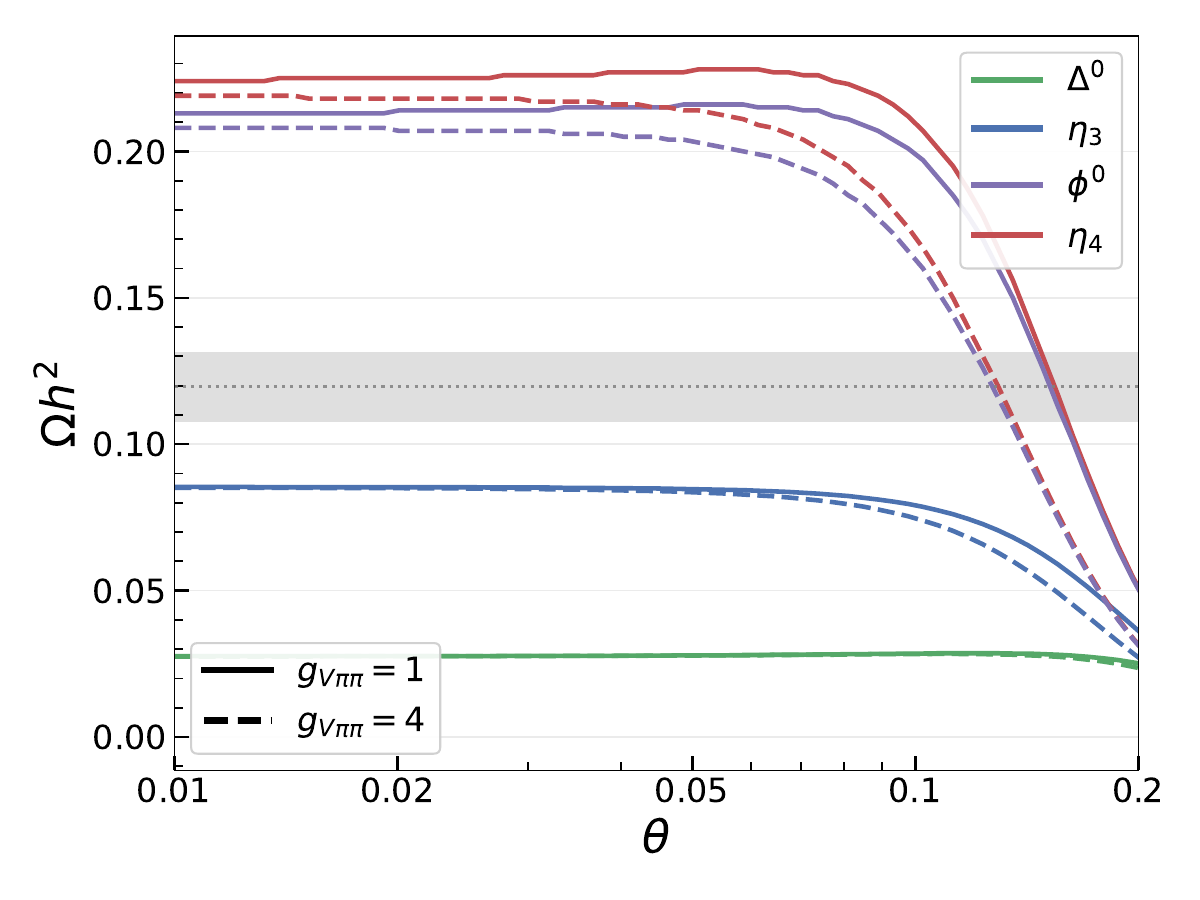}
        \caption{\footnotesize Scan over $\theta$ with $M_\chi = 1~\unit{TeV}$, $\Delta_M = 5~\unit{GeV}$, $M_V = 5~\unit{TeV}$ and $c_i = 0.01$}
        \label{subfig:theta}
    \end{subfigure}
    \hfill
    \begin{subfigure}{0.48\linewidth}
        \centering
        \includegraphics[width=\linewidth]{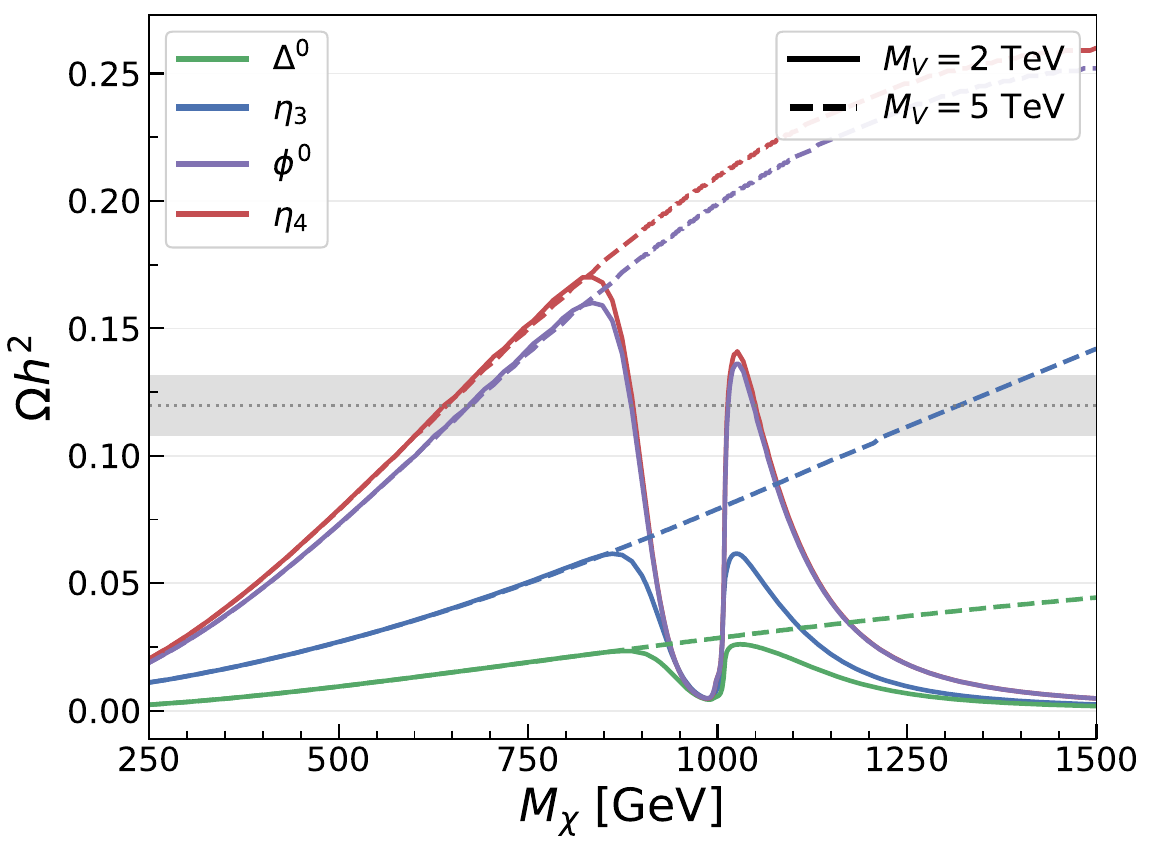}
        \caption{\footnotesize Scan over $M_\chi$ with $\theta = 0.1$, $\Delta_M = 5~\unit{GeV}$, $M_V = 5~\unit{TeV}$ and $c_i = 0.01$}
        \label{subfig:MDM}
    \end{subfigure}

    \vspace{0.25cm}

    \begin{subfigure}{0.48\linewidth}
        \centering
        \includegraphics[width=\linewidth]{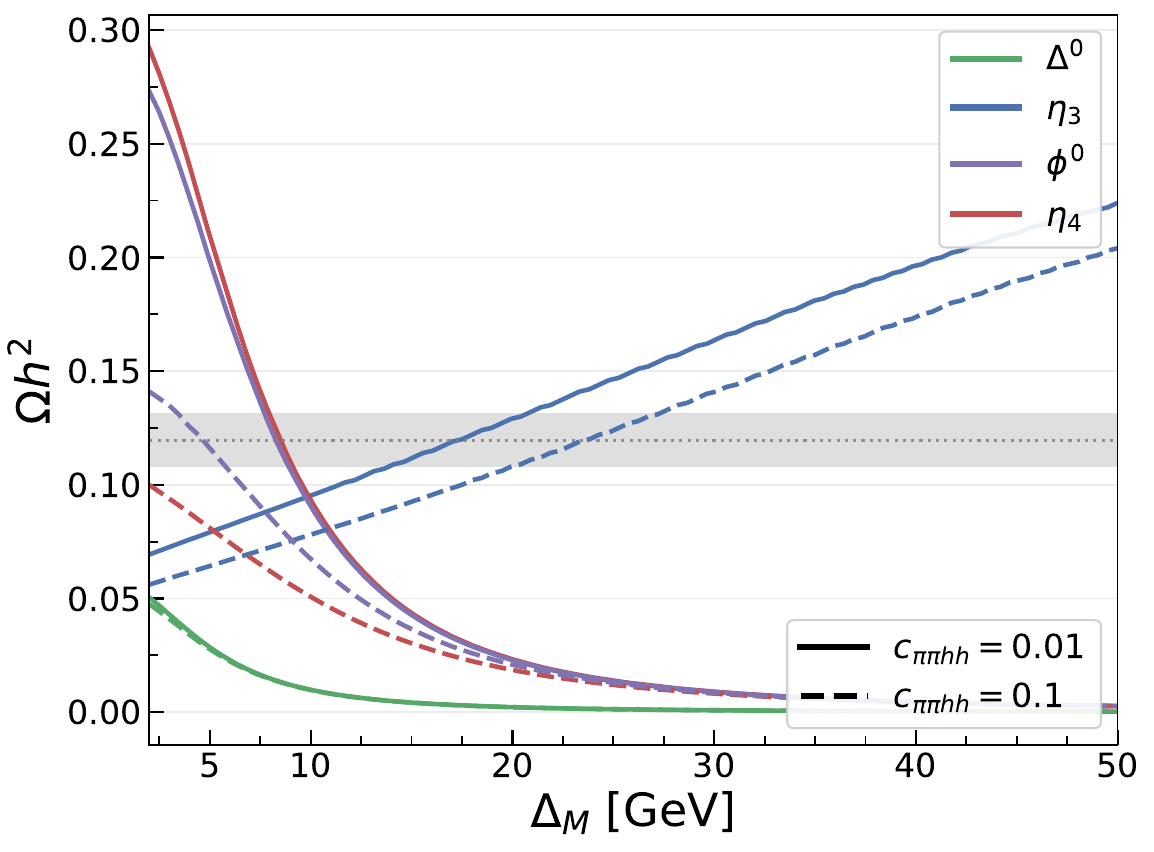}
        \caption{\footnotesize Scan over $\Delta_M$ with $M_\chi = 1~\unit{TeV}$, $\theta = 0.1$, $M_V = 5~\unit{TeV}$ and $c_i = 0.01$}
        \label{subfig:Mdiff}
    \end{subfigure}
    \hfill
    \begin{subfigure}{0.48\linewidth}
        \centering
        \includegraphics[width=\linewidth]{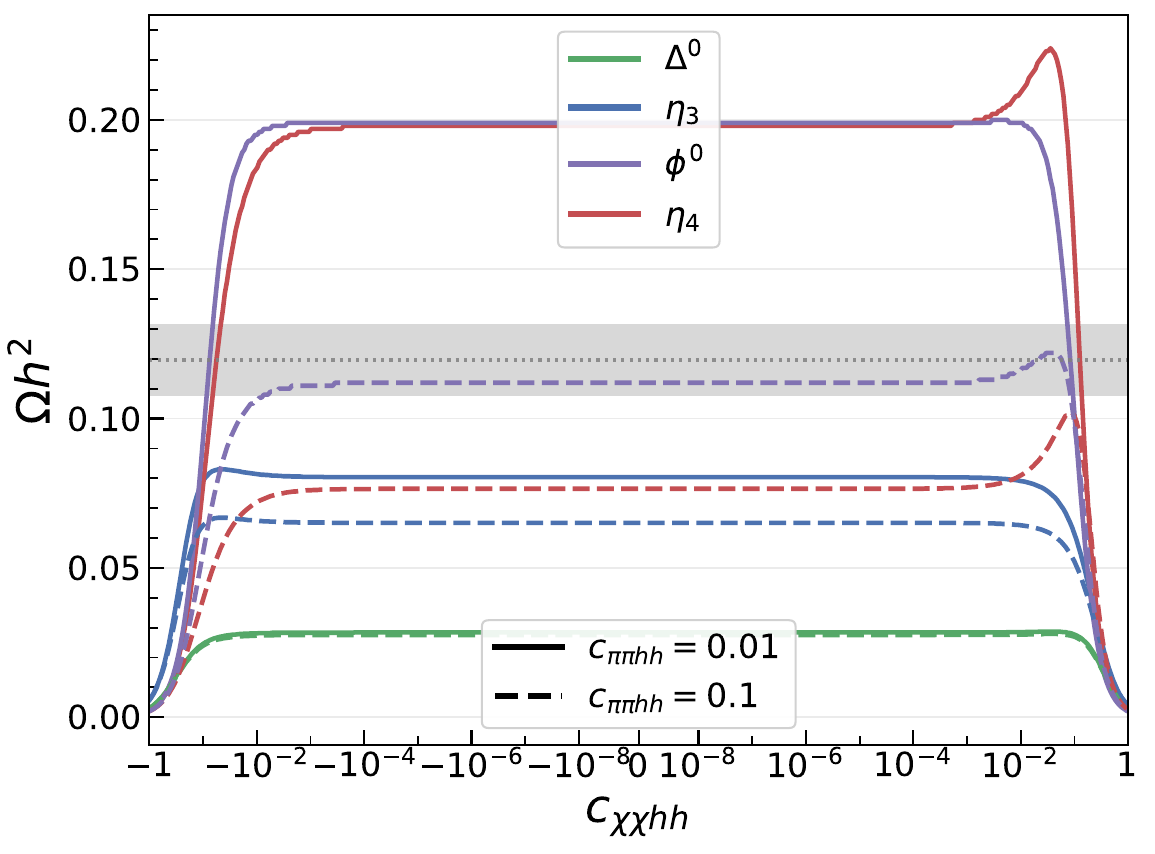}
        \caption{\footnotesize Scan over $c_{\chi\chi hh}$ with $M_\chi = 1~\unit{TeV}$, $\theta = 0.1$, $\Delta_M = 5~\unit{GeV}$ and $M_V = 5~\unit{TeV}$}
        \label{subfig:coupling}
    \end{subfigure}

    \caption{Relic density as a function of the model parameters. The gray band is considered as consistent with the observed DM relic density within the adopted. We have set $g_{V\pi\pi} = 1$, $\tilde{g} = 6$ and $\xi = 1.4$. In (d), $(c_{\chi\chi hh}, c_{\pi\pi hh})$ are respectively $(c_3,c_4)$, $(c_5,c_6)$ or $(c_6,c_5)$ for $\eta_3$, $\eta_4$ or $\phi^0$ DM candidate.}
    \label{fig:parameterscans}
\end{figure}

Before the full parameter scan, we perform a set of one-dimensional parameter variations in order to get an understanding of some basic features. The electroweak input parameters are taken to be $\alpha=1/127.9$, $m_Z=91.1876~\unit{GeV}$, and $G_F=1.1663787\times10^{-5}~\unit{GeV}^{-2}$. We fix the values of model parameters as 
\[
M_\chi=1~\unit{TeV},\;
\Delta M=5~\unit{GeV},\;
M_V=5~\unit{TeV},\;
\xi=1.4,\;
\theta=0.1,\;
g_{V\pi\pi}=1,\; 
\tilde g=6,\; \text{and}\;
c_i=0.01
\]
unless stated otherwise. Starting from this point, we vary one parameter at a time while keeping all other parameters fixed. 
The results are shown in Fig. \ref{fig:parameterscans}. 
Here, we indicate the DM relic density from the Planck Collaboration by a gray dotted line and the assumed uncertainty of ten per-cent by a gray band. 
We see that for this specific parameter point, $\eta_4$ and $\phi^0$ generally have a higher relic density than $\eta_3$ and $\Delta^0$. 
The main reason is that the $\SU(2)_L$ triplet $\Delta$ efficiently annihilates through the SM vector bosons. 
This is also the reason why the relic density is always too low for the $\Delta^0$ DM candidate in the mass range considered. 
In such a scenario one would need an additional contribution to account for the observed relic density. 

We see in Fig.~\ref{subfig:theta} that the relic density is largely insensitive to $\theta$ for small values. However, for $\theta \gtrsim 0.1$ it decreases rapidly with increasing $\theta$. 
This is consistent with the fact that $\theta$ measures the degree of vacuum misalignment and thus the mixing between the composite and EW sectors. 
Moreover, we indicate the importance of the pNGB coupling $g_{V\pi\pi}$ to vectors by setting it to two different values, 1 (solid lines) and 4 (dashed lines). 
Larger values of $g_{V\pi\pi}$ reduce the relic density, with the effect becoming more pronounced with larger $\theta$.

In Fig.~\ref{subfig:MDM}, we vary the mass of the DM candidate and observe that, as expected, the relic density generally increases with $M_\chi$. Different values of $M_V$ have only a minor impact, except in regions where $M_\chi \simeq M_V/2$ and where $M_\chi$ approaches $M_V$. In the former case, the $s$-channel exchange of spin-1 resonances shown in Fig.~\ref{fig:anni} becomes resonant, leading to a significant enhancement of the co-annihilation cross section and thus a reduction of the relic density.
The significant decrease of relic density when $M_\chi$ gets close to $M_V$ is due to the opening of new (co)annihilation channels into one heavy resonance and one SM particle in the final state. 
These additional channels are not listed in Tables \ref{tab:mediatorsEtaDelta} and \ref{tab:mediatorsPhi}, but can be obtained straightforwardly by replacing one of the final-state electroweak gauge bosons with a heavy resonance in $R^0$ or $R^\pm$, while leaving the mediator unchanged. 
The resulting channels increase the effective annihilation cross section and therefore reduce the relic abundance.

The importance of co-annihilation channels is governed by the mass difference $\Delta_M$ between the DM candidate and the additional $\mathbb{Z}_2$-odd pNGBs considered. 
The corresponding dependence of the relic density on $\Delta_M$ is shown in Fig.~\ref{subfig:Mdiff}. And the behaviour differs significantly between the four DM candidates. 
For $\eta_3$ DM, a larger $\Delta_M$ implies a reduced co-annihilation involving $\Delta$, which yields an increase of the relic density. 
For the remaining three candidates, the opposite behaviour is observed: the relic density decreases rapidly as $\Delta_M$ increases. 
For small mass splittings, the co-annihilating heavier $\ZZ$-odd pNGBs become nearly degenerate with the DM candidate, approximately restoring the underlying $\SU(2)_L$ multiplet structure. That is, the $\eta_4$ and $\phi$ states form a doublet, and the $\Delta$ states form a triplet.
In this limit, the $t$-channel exchange of these co-annihilating heavier $\ZZ$-odd pNGB states (topology (h) in Fig.~\ref{fig:anni}) interferes destructively with the contact interaction (topology (i)), leading to a suppressed annihilation cross section into vector bosons and hence an enhanced relic density. 
As the mass splitting increases, the heavier pNGBs decouple from the process, the cancellation becomes less effective, and the relic density decreases.

The effect of the coupling constants in the scalar potential is shown in Fig.~\ref{subfig:coupling}. 
Varying the coupling between the DM candidate and the Higgs field, $c_{\chi\chi hh}$, we observe a behaviour similar to that of $\theta$: insensitivity at small values and a rapid decrease beyond a certain point. The asymmetric behaviour under sign flipping of $c_{\chi\chi hh}$ arises from interference effects in the annihilation channels involving the $\chi$-$\chi$-$h$ vertex. For $\eta_4$ and $\phi^0$, a stronger co-annihilating pNGB--Higgs coupling $c_{\pi\pi hh}$ significantly reduces the relic density. The dependence is weaker for $\eta_3$ and negligible for $\Delta^0$.
We have seen that the Higgs channel, Fig.~\ref{fig:anni} (d)-(f), can be quite important. 
However, it is significantly constrained by direct detection limits, as we shall discuss subsequently.

\subsection{MCMC scans}

The one-dimensional parameter variations discussed in the previous subsection provide only limited insight into the viable regions of the high-dimensional parameter space. 
We therefore perform a Markov Chain Monte Carlo (MCMC) scan based on the Metropolis--Hastings algorithm \cite{Metropolis:1953am,Hastings:1970aa}, in order to get the probability distribution of parameters from our model based on the experimental data. 
Similar explorations of scotogenic models in refs.~\cite{Sarazin:2021nwo,Alvarez:2023dzz} have shown that flavour observables in the lepton/neutrino sector and DM observables constrain complementary sets of parameters\footnote{This was also found in parameter space explorations using AI-assisted parameter scans \cite{deSouza:2025uxb,deSouza:2026jww}.}. 
Thus, we consider here only the DM relic density as well as the direct detection cross sections for DM as observables. 

The scan is performed in an eight-dimensional parameter space,
\begin{equation}
    \Theta = \bigl( M_\chi,\, \Delta_M,\, c_{\chi\chi hh},\, c_{\pi\pi hh},\, \theta,\, g_{V\pi\pi}, \,\tilde{g},\, M_V \bigr),
\end{equation}
where each parameter is restricted to the range, as listed in Table \ref{tab:param_range}. 
Again, $(c_{\chi\chi hh}, c_{\pi\pi hh})$ are
$(c_3,c_4)$, $(c_5,c_6)$ or $(c_6,c_5)$ in the case of $\eta_3$, $\eta_4$ or $\phi^0$ as DM candidate, respectively. We did not find any valid parameter combination for the $\Delta^0$ DM candidate.
Trial points within the allowed range are generated using a multivariate Gaussian proposal distribution with independently tuned step sizes for each parameter. 
We note for completeness, that in principle electroweak precision data can constrain certain combination of parameters as some of the spin-1 resonances mix with the electroweak SM vector bosons. However, there are also additional contributions from top-partners which come with opposite sign \cite{Frandsen:2022xsz,Cai:2022zqu} and which are not relevant for the current study. Thus, we ignore these constraints in the current study.

\begin{table}[t]
\captionsetup{justification=raggedright}
    \centering
    \begin{tabular}{|c|c|}
        \hline
        \textbf{Parameter} & \textbf{Range} \\
        \hline
        $M_\chi$ & [250, 1500]\\
        $\Delta_M$ & [2, 50]\\
        $M_V$ & [2000, 5000]\\
        $\theta$ & [0.01, 0.2]\\
        $g_{V\pi\pi}$ & [1, 4]\\
        $\tilde{g}$ & [1, 10]\\
        $c_{i}$ ($i=3,4,5,6$) & [$10^{-8}$, 1]\\
        \hline
    \end{tabular}
    \caption{Ranges of selected input parameters for the MCMC scan, with all mass parameters given in GeV.}
    \label{tab:param_range}
\end{table}

For each point $\Theta$ in the parameter space, the relic density $\Omega_\Theta$ and direct detection constraints are computed with \texttt{micrOMEGAs}. Instead of imposing a hard cut, we use a Gaussian likelihood,
\begin{equation}
    \mathcal{L}(\Omega_\Theta) \propto 
    \exp\!\left[
    -\frac{(\Omega_\Theta - \Omega_\mathrm{obs})^2}{2\,\sigma_\Omega^2}
    \right],
\end{equation}
where $\Omega_{\rm obs}$ denotes the observed DM relic density and $\sigma_\Omega$ parametrizes theoretical and numerical uncertainties. We take $\sigma_\Omega = 10\% \cdot \Omega_\mathrm{obs}$. A proposed point $\Theta'$ is accepted according to the Metropolis--Hastings criterion,
\begin{equation}\label{eq:MCMCalgorithm}
    u < \min\left(1, \frac{\mathcal{L}(\Omega_{\Theta'})}{\mathcal{L}(\Omega_{\Theta})}\right),
\end{equation}
with $u \in [0,1]$ a uniform random number. Otherwise, the point is rejected.
We do not use the same prescription for the direct detection cross section as the limits implemented in \texttt{micrOMEGAs~6.2} are based on data from LUX-ZEPLIN~\cite{LZ:2022lsv}, XENON1T~\cite{XENON:2018voc}, DarkSide~\cite{DarkSide:2018kuk}, PICO~\cite{PICO:2019vsc}, and CRESST-III~\cite{CRESST:2019jnq}, which have been superseded by LUX-ZEPLIN 2025~\cite{LZ:2024zvo}.
Therefore, we impose the direct detection constraint only afterward.

In this study, we generated a total of $10 \cdot 20000 = 2\cdot 10^5$ points for each MCMC scan to approximate the sampled likelihood distribution. 
Each chain was initialized from a phenomenologically viable point and evolved independently.
\begin{figure}
    \captionsetup{justification=raggedright}
    \centering
    \includegraphics[width=1.0\textwidth]{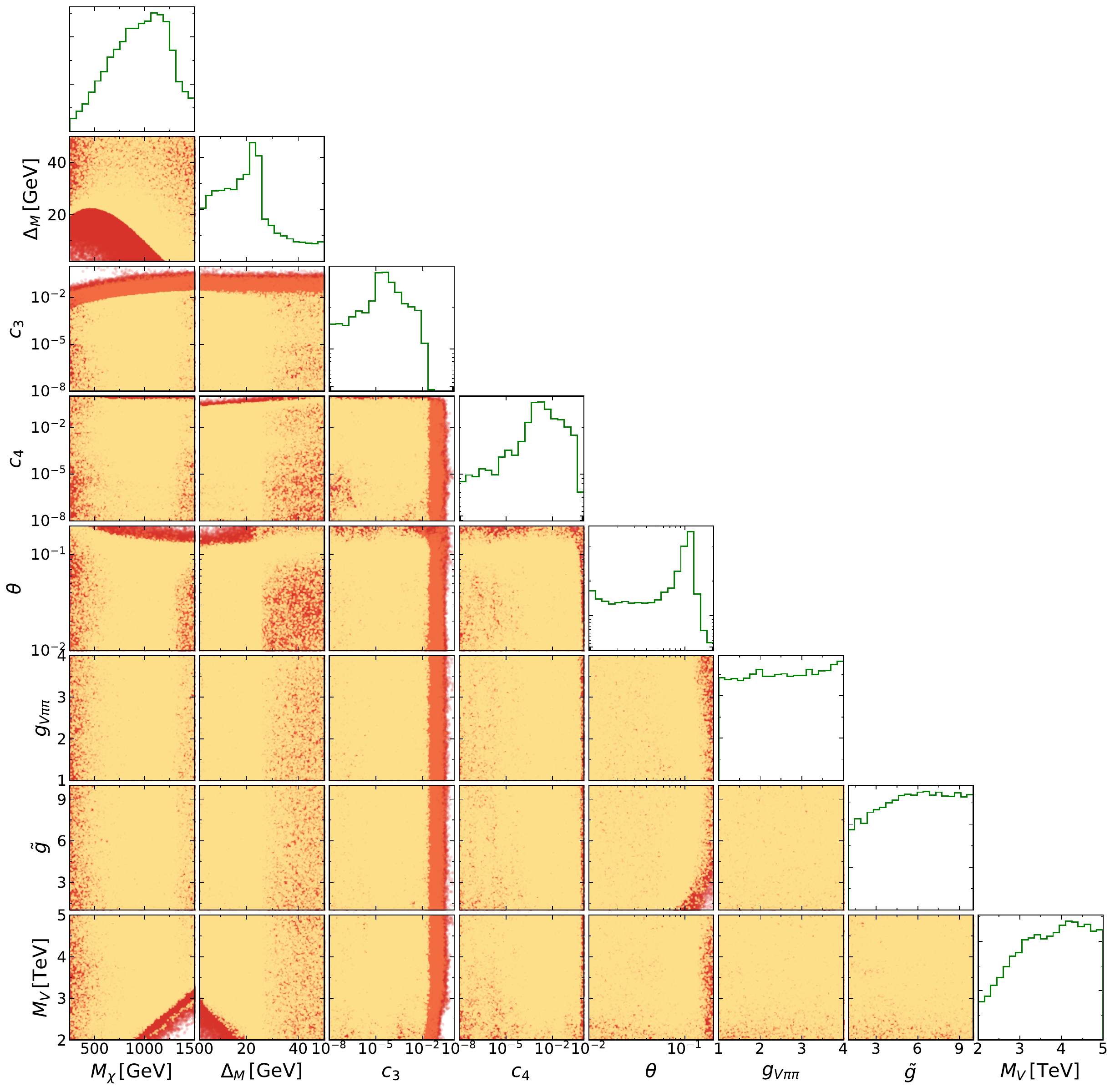}
    \caption{Corner plot of the MCMC scan for the $\eta_3$ DM candidate.}
    \label{fig:eta3_MCMC}
\end{figure}

\begin{figure}
    \captionsetup{justification=raggedright}
    \centering
    \includegraphics[width=1.0\textwidth]{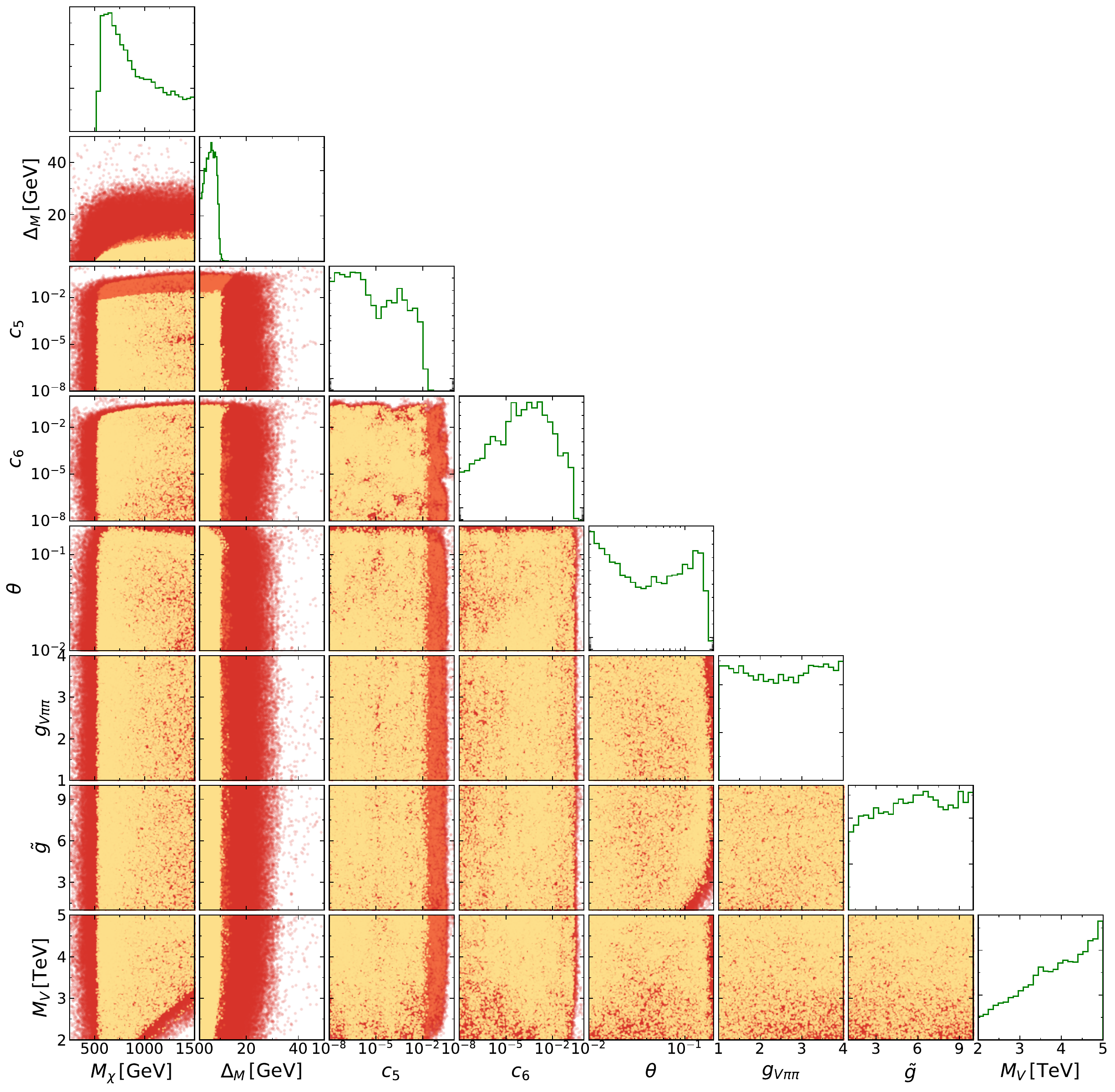}
    \caption{Corner plot of the MCMC scan for the $\eta_4$ DM candidate.}
    \label{fig:eta4_MCMC}
\end{figure}

\begin{figure}
    \captionsetup{justification=raggedright}
    \centering
    \includegraphics[width=1.0\textwidth]{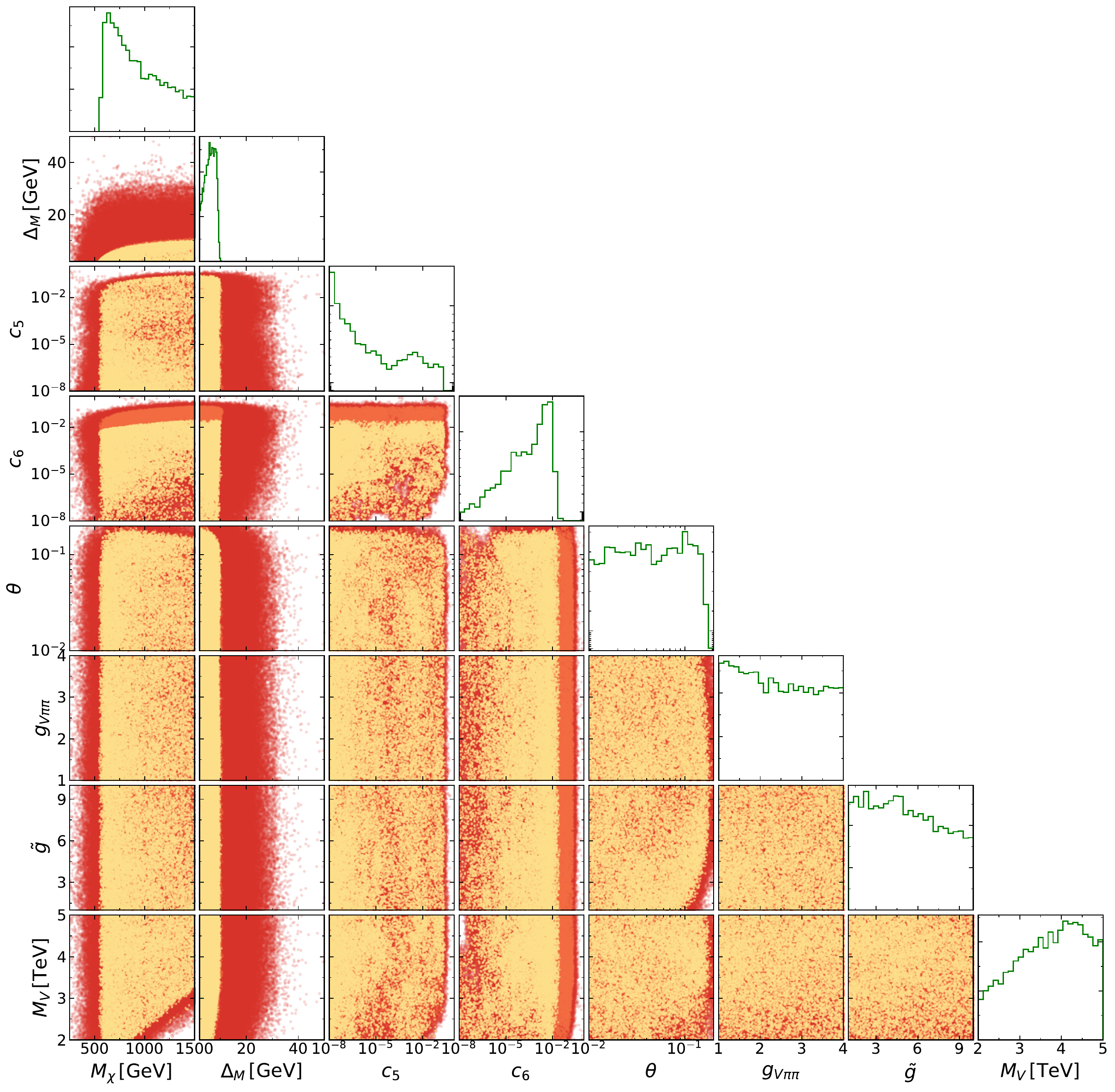}
    \caption{Corner plot of the MCMC scan for the $\phi^0$ DM candidate.}
    \label{fig:phi0_MCMC}
\end{figure}

Corner plots of the MCMC scan for each DM candidate are presented in \cref{fig:eta3_MCMC,fig:eta4_MCMC,fig:phi0_MCMC}, showing the correlations among the model parameters.
Red dots represent all points generated during the MCMC scan, while orange dots denote the points that are accepted according to \cref{eq:MCMCalgorithm} \textit{and} yield the correct relic density within uncertainties. 
The yellow dots correspond to the subset that further satisfy direct detection constraints from LUX-ZEPLIN 2025~\cite{LZ:2024zvo}. 
The diagonal panels display the marginalized one-dimensional distributions of each parameter of the yellow subset. 

The MCMC results reveal several characteristic features of the viable parameter space, which depends on the nature of the DM candidate.
For $\eta_3$ DM, there is an excluded region on the $(M_\chi,\Delta_M)$ plane at small $M_\chi$ and small $\Delta_M$, indicating the large efficiency of co-annihilation processes for lighter $\eta_3$ masses.
The exclusion due to resonance enhancement at $M_\chi \simeq M_V/2$ and the opening of additional annihilation channels when $M_\chi$ approaches $M_V$ is also clearly visible in the $(M_\chi,\, M_V)$ plane. 
The distribution of $\Delta_M$ has a peak around $25~\unit{GeV}$, while the region with simultaneously small $\Delta_M$ and small $M_V$ is not populated.
Too large $\Delta_M$ typically leads to an insufficient co-annihilation rate and therefore to an over-production of DM, which can be seen from the less populated region of large $\Delta_M$, especially in combination with small values of $\theta$, which further reduce the annihilation rate.

The relic density and direct detection constraints have only a mild dependence on the couplings $g_{V\pi\pi}$ and $\tilde g$, whose distributions remain relatively flat across the scanned ranges. 
In contrast, the scan shows a clear preference for large values of the vacuum misalignment angle $\theta$ of about $0.1$. 
The distribution of $M_V$ is also relatively broad, with a slight preference for heavier spin-1 resonances. 
Upper bounds are observed for both $c_{3}$ and $c_{4}$ from the relic density, whereas direct detection constraints push the limit on $c_{3}$ further down to $\mathcal{O}(10^{-2})$, reflecting the limits on scalar-mediated direct DM scattering.

The results for $\eta_4$ and $\phi^0$ DM are qualitatively different. 
In particular, we find a strict upper bound on the mass splitting, $\Delta_M \lesssim 10~\unit{GeV}$. For a smaller mass splitting, destructive interference effects can suppress the annihilation cross section sufficiently to reproduce the observed relic density, as described above.
For $M_\chi$, a sharp lower bound of approximately $500~\unit{GeV}$ is observed, which originates from the requirement of a correct relic density. 
This is a consequence of the electroweak couplings to $Z$ and $W^\pm$ bosons that give rise to very efficient (co-)annihilation cross sections.
The resonant regions, where the relic density would be further strongly reduced, are correspondingly underpopulated.

Although the upper limits on $c_{\chi\chi hh}$ are similar to the case of $\eta_3$, slightly stricter bounds on $c_{\pi\pi hh}$ are obtained in these two scenarios, indicating a more significant impact of co-annihilation in these two cases. 
The distributions of $\tilde g$, $\theta$, and $g_{V\pi\pi}$ remain comparatively flat.

\begin{figure}[h]
\captionsetup{justification=raggedright}
    \centering
    \begin{subfigure}{0.49\textwidth}
        \centering
        \includegraphics[width=\linewidth]{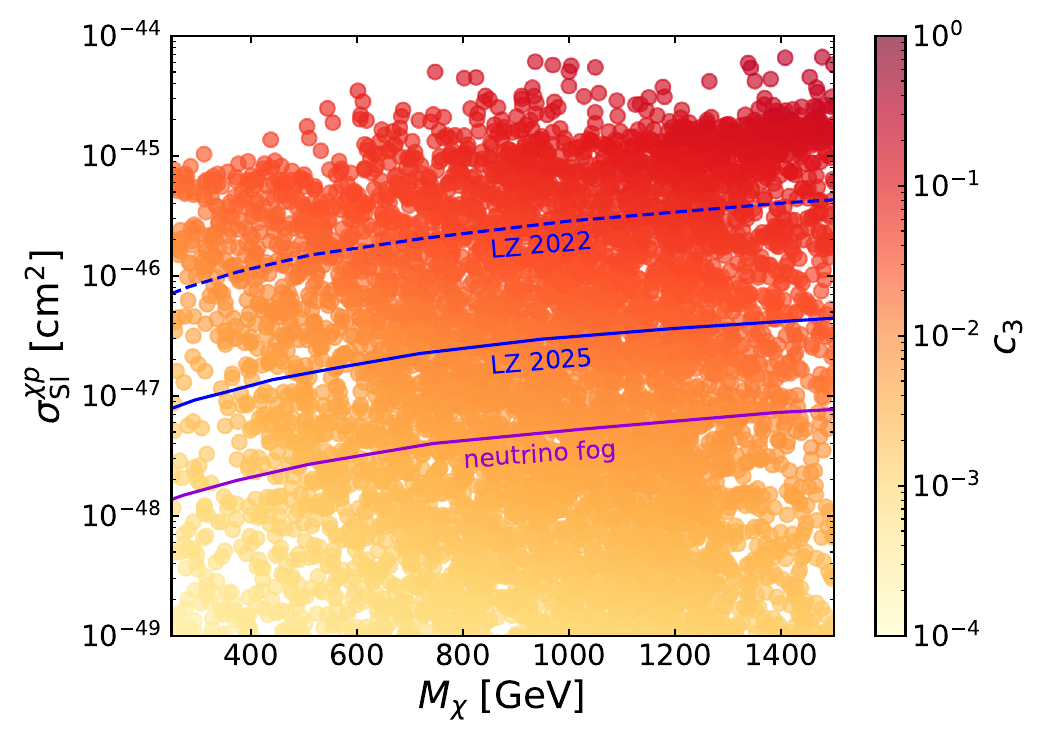}
        \caption{$\eta_3$ DM}
        \label{fig:eta3_DD}
    \end{subfigure}

    \vspace{0.2cm}

    \begin{subfigure}{0.49\textwidth}
        \centering
        \includegraphics[width=\linewidth]{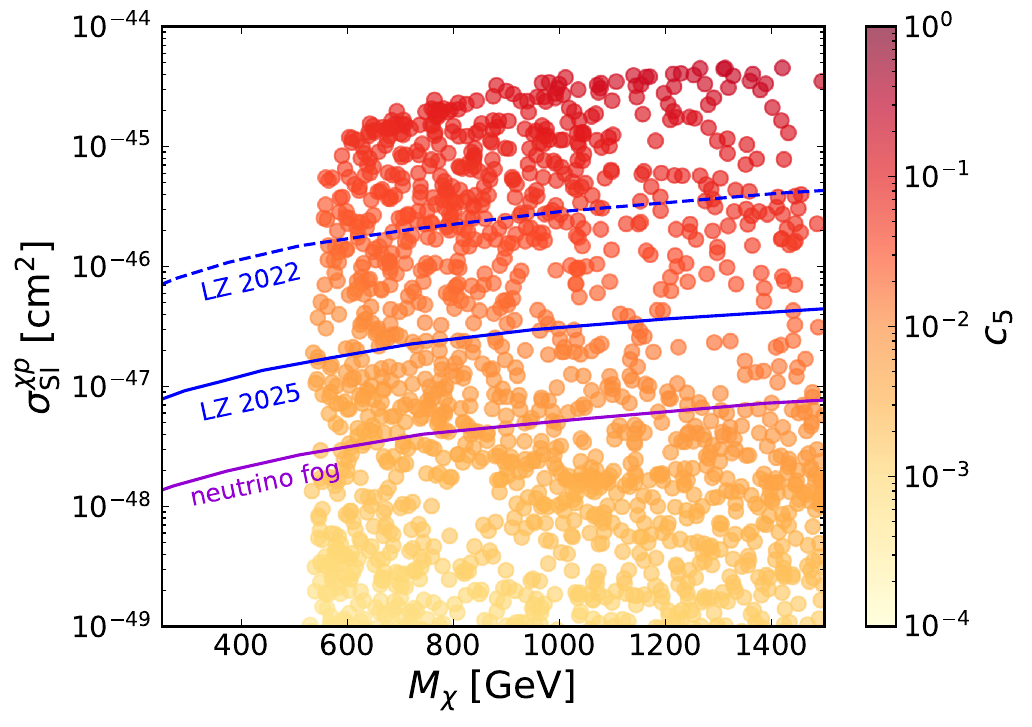}
        \caption{$\eta_4$ DM}
        \label{fig:eta4_DD}
    \end{subfigure}
    \hfill
    \begin{subfigure}{0.49\textwidth}
        \centering
        \includegraphics[width=\linewidth]{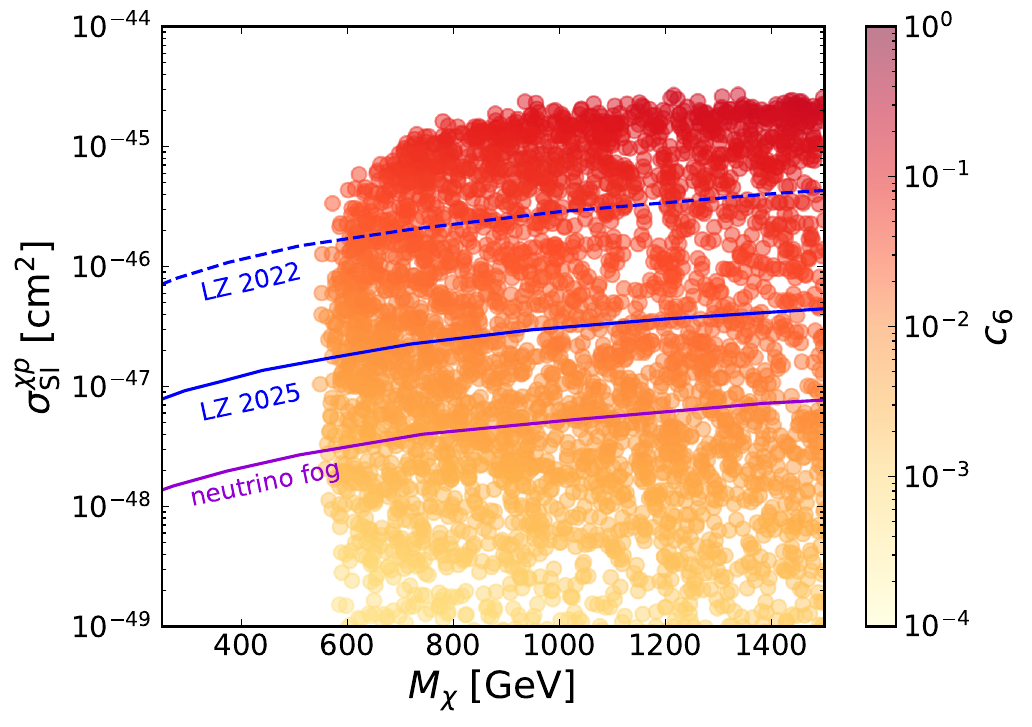}
        \caption{$\phi^0$ DM}
        \label{fig:phi0_DD}
    \end{subfigure}

    \caption{Spin-independent direct detection cross section versus DM mass.}
    \label{fig:DMDD_combined}
\end{figure}

In Fig. \ref{fig:DMDD_combined} we show the spin-independent DM--nucleon scattering cross section $\sigma_{\rm SI}^{\chi p}$ vs DM mass $M_\chi$ for data points from the MCMC scans that give the correct DM relic density within a ten per-cent range. 
The solid line labelled ``LZ 2025'' indicates the latest exclusion curve from LZ 2025~\cite{LZ:2024zvo}. The dashed line shows the bound using the LZ~\cite{LZ:2022lsv} data as implemented in  \texttt{micrOMEGAs} for comparison, demonstrating that the latest results exclude an additional substantial part of the parameter space. Moreover, the figures show that for all three DM scenarios there will always be a sizeable part of the parameter space within the so-called neutrino fog~\cite{OHare:2021utq}.
A clear correlation is observed between $c_{\chi\chi hh}$ and $\sigma_{\rm SI}^{\chi p}$: larger values of $c_{\chi\chi hh}$ generally correspond to larger scattering cross sections, whereas smaller couplings populate the region below the current experimental limits. 
Future direct detection experiments such as DARWIN are expected to reach sensitivities of order $10^{-48}$--$10^{-49}\,\unit{cm}^2$ in the relevant DM mass range \cite{DARWIN:2016hyl}. 
Such sensitivities would probe couplings as small as $c_{\chi\chi hh}\sim10^{-3}$.

\section{A sketch of the LHC phenomenology}
\label{sec:lhc}

The proposed models have served as a basis for several phenomenological studies:
for electroweak pNGBs see \cite{Ferretti:2016upr,Agugliaro:2018vsu,Ruhdorfer:2019utl,Cacciapaglia:2022bax,Flacke:2023eil},
for QCD-coloured pNGBs \cite{Cacciapaglia:2015eqa,Belyaev:2016ftv,Cacciapaglia:2020vyf,Flacke:2025xwl}.
Spin-1 resonances carrying electroweak 
charges are discussed in \cite{BuarqueFranzosi:2016ooy,Caliri:2024jdk} and those with QCD charges in \cite{Cacciapaglia:2024wdn}.
Top-partners with non-standard decays are discussed in \cite{Bizot:2018tds,Xie:2019gya,Cacciapaglia:2019zmj,Banerjee:2022izw,Banerjee:2024zvg} and those containing also colour octets and sextets in \cite{Cacciapaglia:2021uqh,Cacciapaglia:2026jlv}.

An obvious implication of the $\ZZ$ symmetry in this class of models is the presence of events with large missing transverse momentum, similar to those in supersymmetric models. 
This would show up for example in the production of two top quarks in association with two pNGBs as discussed in \cite{Haisch:2021ugv}.
Another possibility is the pair production of the additional pNGBs via Drell-Yan processes: 
\begin{itemize}
 \item The $\ZZ$-even pNGBs will lead to either final states with several third generation quarks or multi vector bosons depending on the precise scenario \cite{Flacke:2025xwl}.
 \item The $\ZZ$-odd pNGBs will lead to events with large missing transverse energy and soft leptons and jets from their cascade decays.
\end{itemize}

In case of spin-1 resonances, the current bounds are already between 1.5 and 4.5 TeV depending on the model details. 
This clearly implies that, the pair production of these states at the LHC will only play a minor role. 
However, there could still be the possibility of an associated production of a spin-1 resonance with either a Higgs boson or electroweak vector boson:
\begin{itemize}
 \item A $\ZZ$-even spin-1 resonance could be produced together with either a Higgs boson, a $W$- or a $Z$-boson. This will give rise to hard multi-lepton and/or quark (jet) final states.
 \item A $\ZZ$-odd spin-1 resonance can only be produced together with a $\ZZ$-even pNGB. The spin-1 resonance will in this case decay dominantly into a $\ZZ$-odd pNGB plus either a Higgs boson, a $W$- or a $Z$-boson. This will lead to hard jets/leptons plus large missing transverse energy.
\end{itemize}
Note that one expects the spin-1 resonances to have masses of order $M_V$.
At a prospective 100 TeV $pp$-collider one could pair produce the spin-1 resonances, again giving rise to both types of signatures: hard jets/leptons combined with/without large missing transverse momentum.

A similar situation occurs for the  top-partners.
Also, this sector will contain both $\ZZ$-odd and $\ZZ$-even types of states.  
The $\ZZ$-odd ones will have signatures very much like the scalar partners of the third generation quarks in supersymmetric models, see for example \cite{Bartl:1994bu,Porod:1996at,Bartl:1997yi,Bartl:1998xk,Porod:1998yp}.
We note for completeness that, missing energy signals for top-partners can also occur in case of an accidental `baryon' quantum number \cite{Cacciapaglia:2021uqh}. 
The characteristic feature of the scotogenic models is that the missing energy signatures will be present in every sector.

\section{Conclusions and outlook}
\label{sec:onclusions}

We have investigated DM phenomenology in a specific composite Higgs model with a fermionic UV completion that leads to an $\SU(6)/\Sp(6)$ breaking pattern at low-energy. 
A stable $\mathbb{Z}_2$-odd sector containing four neutral pNGB states naturally appears. 
We have found that, for three out of those four, viable regions in parameter space exist in which the observed DM relic density can be explained.

We have used MCMC scans to identify the parameter space consistent with the observed relic density and current DM direct detection limits.
Although we have made some simplifying assumptions to reduce the number of parameters, the overall features are expected to be robust even when we give up these assumptions about the underlying couplings and mass spectrum.

An important finding of our analysis is that spin-1 resonances have a significant impact on the DM relic density in certain regions of parameter space.
This naturally raises the question on the importance of other resonances predicted by this class of models, in particular the top-partners which are required in models with partial compositeness. 
Some of the top-partners will be $\ZZ$-odd, implying that they can occur in t-channel processes when the dark matter particles annihilate to third generation quarks.
This clearly induces some more model dependence and we leave the corresponding case studies for future investigations.

Last but not least, we want to stress that this class of models has a fascinating LHC phenomenology. The production of $\ZZ$-even particles will lead to high energetic leptons and jets typical for composite Higgs models. In contrast, the production of $\ZZ$-odd particles will lead to signatures with large missing transverse momentum which are typical for supersymmetric models. 

\begin{acknowledgments}
We thank G.~Cacciapaglia, R.~Caliri, J.~Hadlik and M.~Kunkel for useful discussions. 
This work has been supported by DFG  GRK 2994 and project nr. PO-1337/12-1.
\end{acknowledgments}

\appendix

\section{Technical aspects of the considered model}
\label{app:technical_aspects}

The model considered in this work is based on the coset $\SU(6)/\Sp(6)$. 
The basic structure of this coset has been discussed in \cite{Cai:2018tet,Cacciapaglia:2020psm}. 
We summarize here technical details which were used in obtaining the necessary interactions.

\subsection{Underlying fermionic theory and $\SU(2)_L\times \SU(2)_R$ embedding}
\label{sec:structure}

In the underlying UV theory, six hyper-fermions transform as the fundamental $\mathbf{6}$ of an $\SU(6)$ global symmetry group. This group is spontaneously broken to $\Sp(6)$ if the hyper-fermions are in a pseudo-real representation of the confining group $G_{HC}$. 
This gives rise to a condensate of the form
\begin{equation}
    \langle \psi_i^{\alpha,a} \psi_j^{\beta,b} \rangle \epsilon_{\alpha\beta}\epsilon_{ab} \sim \Sigma_{ij},
\end{equation}
which is antisymmetric with respect to the $\SU(6)$ indices $i$ and $j$. Here, $\alpha,\;\beta$ and $a,\;b$ are respectively spinor and $G_{HC}$ indices.
Thus, the corresponding vacuum structure has the form
\begin{equation}\label{eq:sigma0}
    \langle\psi\psi\rangle \sim \Sigma_0 = \begin{pmatrix}
        i \sigma_2 & 0 & 0\\
        0 & -i \sigma_2 & 0\\
        0 & 0 & i \sigma_2
    \end{pmatrix},
\end{equation}
which is left invariant by $\Sp(6)$ transformations
\begin{equation}
    \Sigma_0 \to U \cdot\Sigma_0\cdot U^T.
\end{equation}
The 21 unbroken generators $S_i$ of $\Sp(6)$ fulfil the condition 
\begin{equation}\label{eq:unbrokengen}
    S_i\cdot\Sigma_0 + \Sigma_0\cdot S_i^T = 0\,,
\end{equation}
and the 14 broken generators $X_j$ of $\SU(6)$
\begin{equation}
    X_j\cdot\Sigma_0 - \Sigma_0\cdot X_j^T = 0 \,.
\end{equation}

The $\Sp(6)$ group contains three commuting $\SU(2)$ subgroups, two of which can be identified with the custodial group $\SU(2)_L \times \SU(2)_R$.
We choose the basis as indicated in \cref{tab:embedding}. The $\SU(2)_L$ subgroup is embedded in $\Sp(6)$ with the generators
\begin{equation}
    T^i_L = \frac{1}{2} \begin{pmatrix}
    \sigma^i & 0 & 0\\
    0 & 0 & 0\\
    0 & 0 & \sigma^i
\end{pmatrix} 
\end{equation} 
 and $\SU(2)_R$ with the generators
\begin{equation}
    T^i_R = \frac{1}{2} \begin{pmatrix}
    0 & 0 & 0\\
    0 & -\sigma^T_i & 0\\
    0 & 0 & 0
\end{pmatrix} \,.
\end{equation}
We note for completeness that the different normalisation of the generators is a consequence of the fact
that we have a direct sum of two $\SU(2)_L$ doublets in \cref{tab:embedding}.

A property of this model is that a discrete $\ZZ$ symmetry can be embedded in the $\Sp(6)$ subgroup: 
\begin{equation}\label{eq:Z2}
    P = \exp{(2i\pi S_9)} = \begin{pmatrix}
        \mathbb{1}_4 & 0 \\
         0 & - \mathbb{1}_2
    \end{pmatrix} \quad \text{ with } \quad
 S_9 = \frac{1}{2}\begin{pmatrix}
    0 & 0 & 0\\
    0 & 0 & 0\\
    0 & 0 & \sigma_3
    \end{pmatrix} \,.    
\end{equation}
$S_9$ is one of the $\Sp(6)$ generators.
The first two doublets of \cref{tab:embedding}
are even with respect to this $\ZZ$ whereas the
third one is odd.

\subsection{pNGBs and vacuum misalignment}

Explicit symmetry breaking, e.g.~through gauging the electroweak interactions or generating the top-mass via partial compositeness, generates a potential for the pNGBs, resulting in the spontaneous breaking of the electroweak symmetry. 
This leads to the so-called vacuum misalignment which rotates the vacuum $\Sigma_0$ to the true vacuum $\Sigma_\theta$. It can be shown, see e.g.~\cite{Cai:2018tet,Caliri:2024jdk} and refs.~therein, that vacuum misalignment can be described in terms of a single parameter $\theta$ if only one pNGB, which is identified with the Higgs boson, causes electroweak symmetry breaking. 
The true vacuum $\Sigma_\theta$ can be expressed as
\begin{equation}
 \Sigma_\theta = \Omega(\theta)\Sigma_0\Omega^T(\theta)= \begin{pmatrix}
            i\sigma_2 \cos{\theta} & \mathbb{1}_2 \sin{\theta} & 0\\
            -\mathbb{1}_2 \sin{\theta} & -i\sigma_2 \cos{\theta} & 0 \\
            0 & 0 & i\sigma_2
            \end{pmatrix},
\end{equation}
using the convention in \cite{Cacciapaglia:2020psm} with \begin{equation}
 \Omega(\theta) = \exp{\sqrt{2}i\theta X_h} \text{ and }
 X_h = \frac{1}{2\sqrt{2}}\begin{pmatrix}
        0 & \sigma_2 & 0 \\
        \sigma_2 & 0 & 0 \\
        0 & 0 & 0
    \end{pmatrix} \,.
\end{equation}

The Goldstone matrix defined around the true vacuum is given by
\begin{equation}
    U(\tilde{\Pi}) = \exp\left(i\frac{\sqrt{2}}{f_\pi}\sum_I \pi_I \tilde{X}_I \right) = \exp(i\frac{\sqrt{2}}{f_\pi}\tilde{\Pi}).
\end{equation}
with rotated generators obtained by \[
\tilde{S}^a = \Omega(\theta)S^a\Omega^\dagger(\theta),\hspace{0.3cm} \tilde{X}^I = \Omega(\theta)X^I\Omega^\dagger(\theta).\]
The decay constant $f_\pi$ is related to the misalignment angle by\begin{equation}
    f_{\pi}\sin{\theta} = v_{\rm SM} = 246~\unit{\GeV}. \,
\end{equation}
 The pNGB matrix $\tilde \Pi$ reads as 
\begin{equation}\label{eq:pNGBs}
   \tilde  \Pi = \frac{1}{2} \Omega(\theta) \begin{pmatrix}
    \frac{1}{\sqrt{2}}\eta_1 + \frac{1}{\sqrt{6}}\eta_2 & H & \Delta+\frac{1}{\sqrt{2}}\eta_3\\
    H^\dagger & -\frac{1}{\sqrt{2}}\eta_1 + \frac{1}{\sqrt{6}}\eta_2 & \Phi^\dagger\\
    \Delta^\dagger+\frac{1}{\sqrt{2}}\eta_3 & \Phi & -\sqrt{\frac{2}{3}}\eta_2
    \end{pmatrix} \Omega^\dagger(\theta)\, .
\end{equation}
The pNGB states contain two $\SU(2)_L\times \SU(2)_R$ bi-doublets, $H$ and $\Phi$, which split into two triplets, $\varphi$ and $\phi$, and two singlets, $h$ and $\eta_4$ due to electroweak symmetry breaking. 
$\varphi$ are the would-be Goldstone bosons forming the longitudinal components of the vector bosons.
$\Delta$ is a triplet under $\text{SU(2)}_{L}$, while the $\eta$'s are singlets. 
Last but not least, note that the $\mathbb{Z}_2$ symmetry generated by $S_9$ is preserved by the true vacuum
\begin{align}
    \tilde{S}_9 = S_9 \,.
\end{align}

\subsection{Hidden symmetry approach and spin-1 resonances}
\label{app:hidden_symmetry}

We use the hidden gauge symmetry formalism~\cite{Bando:1987br} to include
spin-1 resonances in our model. 
We outline here the main steps and refer to ref.~\cite{BuarqueFranzosi:2016ooy} for further details.
The basic idea of this approach is as follows: 
one extends the global flavour group $\SU(6)$ to the direct product of two copies, $\SU(6)_0\times \SU(6)_1$. 
$\SU(6)_0$ corresponds to the usual global flavour group in which the electroweak gauge group is embedded, while $\SU(6)_1$ is fully gauged with initially massless gauge bosons. 
The product group $\SU(6)_0\times \SU(6)_1$ is spontaneously broken to $\Sp(6)_0\times \Sp(6)_1$ due to the condensate of the hyper-fermions.  
The corresponding $\SU(6)_1$ gauge fields split into a vector part $\mathcal{V}_\mu$ and an axial part $\mathcal{A}_\mu$.
This spontaneous symmetry breaking leads to two sets of pNGB fields $\pi_i$ ($i=0,1$).
One linear combination of these pNGBs results in physical states while the other forms the longitudinal components of the axial vector bosons $\mathcal{A}_\mu$.
The breaking of $\Sp(6)_0\times \Sp(6)_1$ to its diagonal subgroup leads to would-be Goldstone bosons which form the longitudinal components of the vector states $\mathcal{V}_\mu$.

On the technical side, we introduce two Goldstone matrices defined around the true vacuum\begin{equation}
    U_i = \exp\left(\frac{i\sqrt{2}}{f_i} \pi_i^I \tilde{X}_I\right), \quad i = 0,1,
\end{equation}
each containing 14 pNGBs. 
Here and below we denote the broken generators by $\tilde{X}_I$ and the unbroken ones by $\tilde{S}_i$. 
The Maurer-Cartan form for each factor reads as
\begin{equation}
    \Omega_{i,\mu} = \text{i}U_i^\dagger D_\mu U_i,
\end{equation}
with 
\begin{align}\label{eq:cov_der0}
    D_\mu U_0 &= \left( \partial_\mu - \text{i}\hat{g}\tilde{W}^i_\mu T_L^i - \text{i}\hat{g}^\prime B_\mu T_R^3 \right) U_0, \\ \label{eq:cov_der1}
    D_\mu U_1 &= \left( \partial_\mu - \text{i}\tilde{g}\mathcal{V}^a_\mu \tilde{S}_a - \text{i}\tilde{g}\mathcal{A}^I_\mu \tilde{X}_I\right) U_1.
\end{align} 
The hats on the couplings in \cref{eq:cov_der0} indicate that these are not yet the usual electroweak gauge couplings of the SM, as we will see below.
We split the Maurer-Cartan forms into $d$- and $e$-symbols, corresponding to broken and unbroken generators, which are part of the basic ingredients of the CCWZ Lagrangian:
\begin{align}
    d_{i,\mu} & =  2 \Tr(\Omega_{i,\mu}\tilde{X}_I)\tilde{X}_I,
    \\
    e_{i,\mu} & =  2 \Tr(\Omega_{i,\mu}\tilde{S}_a)\tilde{S}_a.   
\end{align}
The factor of $2$ is due to our normalization of the generators $\Tr{T^AT^B} = \frac{1}{2}\delta^{AB}$. 
The product group $\Sp(6)_0\times \Sp(6)_1$ is spontaneously broken to its diagonal part $\Sp(6)$ by introducing a scalar multiplet
\begin{equation}
    K = \exp\left(\frac{i}{f_K} k^a \tilde{S}_a\right),
\end{equation}
giving rise to 21 would-be Goldstone bosons for the longitudinal components of the vector fields $\mathcal{V}_\mu$. 
At leading order, the Lagrangian reads as~\cite{BuarqueFranzosi:2016ooy}
\begin{equation}\label{eq:Lag}
\begin{aligned}
    \mathcal{L} = & -\frac{1}{2}\Tr\mathcal{F}_{\mu\nu}\mathcal{F}^{\mu\nu}
    -\frac{1}{2}\Tr\mathbf{W}_{\mu\nu}\mathbf{W}^{\mu\nu}
    -\frac{1}{4} \mathbf{B}_{\mu\nu}\mathbf{B}^{\mu\nu}\\
    & + \frac{f_0^2}{2}\Tr d_{0\mu}d_0^\mu
    + \frac{f_1^2}{2}\Tr d_{1\mu}d_1^\mu
    + rf_1^2\Tr d_{0\mu}Kd_1^\mu K^\dagger
    + \frac{f_K^2}{2}\Tr D^{\mu}K(D_\mu K)^\dagger,
\end{aligned}
\end{equation}
with
\begin{equation}
    D_\mu K = \partial_\mu K - ie_{0,\mu}K + iKe_{1,\mu}\,.
\end{equation} 
And $\mathcal{F}_{\mu\nu}$ is the field strength tensor of spin-1 resonances, and $\mathbf{W}_{\mu\nu}$ and $\mathbf{B}_{\mu\nu}$ are the $\SU(2)_L$ and $\U(1)_Y$ field strength tensors.

In this approach, the decay constant of the physical pNGBs is given by
\begin{equation}\label{eq:fpi}
    f_\pi = \sqrt{f_0^2 - r^2f_1^2} = v_{\text{EW}}/\sin{\theta}.
\end{equation}
The states which do not mix with the electroweak gauge bosons have universal masses
\begin{equation}\label{eq:resonance_masses}
    M_A^2 = \frac{f_1^2\tilde{g}^2}{2},\qquad M_V^2 = \frac{f_K^2\tilde{g}^2}{2} \,.
\end{equation}
Inspecting  \cref{eq:Lag}, one finds that the spin-1 resonances, $a_{1\mu}$, $v_{1\mu}$, $v_{2\mu}$ and $v_{3\mu}$ (see \cref{tab:spectrum}), mix with the electroweak gauge bosons $\tilde{W}_\mu$ and $B_\mu$.
In the basis ($\tilde{W}_\mu^+$, $a_{1\mu}^+$, $v_{1\mu}^+$, $v_{2\mu}^+$, $v_{3\mu}^+$) the mass matrix for the charged states reads
\begin{equation}\label{eq:Cmass}
    \mathcal{M}_C^2 = \begin{pmatrix}
        \frac{\hat{g}^2M_V^2(2+w\text{s}^2_\theta)}{\tilde{g}^2} & -\hat{g}\frac{rM_A^2\text{s}_\theta}{\sqrt{2}\tilde{g}} & -\hat{g}\frac{M_V^2}{\sqrt{2}\tilde{g}} & -\hat{g}\frac{M_V^2 c_\theta}{\sqrt{2}\tilde{g}} & -\hat{g}\frac{M_V^2}{\tilde{g}}\\
        
        -\hat{g}\frac{rM_A^2\text{s}_\theta}{\sqrt{2}\tilde{g}} & M_A^2 & 0 & 0 & 0 \\
        
        -\hat{g}\frac{M_V^2}{\sqrt{2}\tilde{g}} & 0 & M_V^2 & 0 & 0 \\
        
        -\hat{g}\frac{M_V^2 c_\theta}{\sqrt{2}\tilde{g}} & 0 & 0 & M_V^2 & 0 \\
        
        -\hat{g}\frac{M_V^2}{\tilde{g}} & 0 & 0 & 0 & M_V^2        
    \end{pmatrix},
\end{equation}
with $w = (f_0^2/f_K^2-1)/2$. 
It is clear from this mass matrix that only 
a linear combination of $v_{1\mu}^+$, $v_{2\mu}^+$ and $v_{3\mu}^+$ mixes with $\tilde{W}_\mu^+$. 
Moreover, we see that the mixing of $\tilde{W}_\mu^+$ with $a_{1\mu}^+$ vanishes in the limit $\theta\to 0$. 

In the neutral sector, the mass matrix in
the basis ($B_\mu$, $\tilde{W}_\mu^3$, $a_{1\mu}^0$, $v_{1\mu}^0$, $v_{2\mu}^0$, $v_{3\mu}^0$) is given by
\begin{equation}\label{eq:Nmass}
    \mathcal{M}_N^2 = \begin{pmatrix}
        \frac{\hat{g}'^2M_V^2(1+w\text{s}^2_\theta)}{\tilde{g}^2} & -\frac{\hat{g}\hat{g}'M_V^2w\text{s}^2_\theta}{\tilde{g}^2} & -\hat{g}'\frac{rM_A^2\text{s}_\theta}{\sqrt{2}\tilde{g}}& -\hat{g}'\frac{M_V^2}{\sqrt{2}\tilde{g}} & \hat{g}'\frac{M_V^2 c_\theta}{\sqrt{2}\tilde{g}} & 0\\
        
        -\frac{\hat{g}\hat{g}'M_V^2w\text{s}^2_\theta}{\tilde{g}^2} & \frac{\hat{g}^2M_V^2(2+w\text{s}^2_\theta)}{\tilde{g}^2}& \hat{g}\frac{rM_A^2\text{s}_\theta}{\sqrt{2}\tilde{g}}&-\hat{g}\frac{M_V^2}{\sqrt{2}\tilde{g}}& -\hat{g}\frac{M_V^2 c_\theta}{\sqrt{2}\tilde{g}} & -\hat{g}\frac{M_V^2}{\tilde{g}}\\
        
        -\hat{g}'\frac{rM_A^2\text{s}_\theta}{\sqrt{2}\tilde{g}} & \hat{g}\frac{rM_A^2\text{s}_\theta}{\sqrt{2}\tilde{g}} & M_A^2 & 0 & 0 & 0\\
        
        -\hat{g}'\frac{M_V^2}{\sqrt{2}\tilde{g}} & -\hat{g}\frac{M_V^2}{\sqrt{2}\tilde{g}} & 0 & M_V^2 & 0 & 0 \\
        
        \hat{g}'\frac{M_V^2 c_\theta}{\sqrt{2}\tilde{g}} & -\hat{g}\frac{M_V^2 c_\theta}{\sqrt{2}\tilde{g}} & 0 & 0 & M_V^2 & 0\\
        
        0 & -\hat{g}\frac{M_V^2}{\tilde{g}} & 0 & 0 & 0 & M_V^2      
    \end{pmatrix}.
\end{equation}
Here, two linear combinations of $v_{1\mu}^0$, $v_{2\mu}^0$ and $v_{3\mu}^0$ mix with photon and $Z$-boson, while the mixing with $a_{1\mu}^0$  vanishes for $\theta \to 0$.

Each mass matrix is diagonalized by an orthogonal rotation matrix, denoted by $\mathcal{C}$ and $\mathcal{N}$ for the charged and neutral vectors, respectively.
The gauge eigenstates and the mass eigenstates are related by
\begin{equation}
    \begin{pmatrix}
    \tilde{W}_\mu^+ \\
    a_{1\mu}^+ \\
    v_{1\mu}^+ \\
    v_{2\mu}^+\\
    v_{3\mu}^+
    \end{pmatrix}
    =\mathcal{C}
    \begin{pmatrix}
    W_\mu^+ \\
    A_{1\mu}^+ \\
    V_{1\mu}^+ \\
    V_{2\mu}^+\\
    V_{3\mu}^+
    \end{pmatrix}
    =\mathcal{C}R^+_\mu, \qquad
    \begin{pmatrix}
    B_\mu\\
    \tilde{W}_\mu^3 \\
    a_{1\mu}^0 \\
    v_{1\mu}^0 \\
    v_{2\mu}^0\\
    v_{3\mu}^0
    \end{pmatrix}
    = \mathcal{N}
    \begin{pmatrix}
    A_\mu \\
    Z_\mu \\
    A_{1\mu}^0\\
    V_{1\mu}^0 \\
    V_{2\mu}^0\\
    V_{3\mu}^0
    \end{pmatrix}=\mathcal{N}R^0_\mu \,.
\end{equation}
In case of the photon, the corresponding elements of ${\cal N}$ can be found analytically and one obtains
\begin{equation}\label{eq:photon}
    A_\mu = \frac{e}{\hat{g}^\prime} B_\mu + \frac{e}{\hat{g}}\tilde{W}_\mu^3 + \frac{\sqrt{2}e}{\tilde{g}}v_{1,\mu}^0 + \frac{e}{\tilde{g}}v_{3,\mu}^0,
\end{equation}
with
\begin{equation}\label{eq:e}
    \frac{1}{e^2} = \frac{1}{\hat{g}'^2} + \frac{1}{\hat{g}^2} + \frac{3}{\tilde{g}^2}.
\end{equation}

\subsection{Relevant interactions}
The Lagrangian in \cref{eq:Lag} contains interactions of pNGBs among themselves, involving derivatives, as well as with the heavy spin-1 resonances. Those interactions involving DM candidates can play a role in the DM phenomenology. In this section, we categorize the relevant interaction types that contribute to DM phenomenology, in particular relic density and direct detection, as discussed below.

We expand the Goldstone matrices up to $\mathcal{O}(1/f^3_i)$:
\begin{equation}
    U_i \approx \mathbb{I}_6 + \frac{i\sqrt{2}}{f_i}\tilde{\Pi}_i + \frac{1}{2!}\left(\frac{i\sqrt{2}}{f_i}\right)^2\tilde{\Pi}_i^2 + \frac{1}{3!}\left(\frac{i\sqrt{2}}{f_i}\right)^3 \tilde{\Pi}_i^3 + \mathcal{O}\left(1/f^4_i\right),
\end{equation}
then the Maurer-Cartan forms become
\begin{align*}
    \text{i}U_i^\dagger D_\mu U_i & \approx  i\left(\mathbb{1}_6 - \frac{i\sqrt{2}}{f_i} \tilde{\Pi}_i - \frac{1}{f_i^2}\tilde{\Pi}_i^2 + \frac{i\sqrt{2}}{3 f_i^3}\tilde{\Pi}_i^3\right)\left(\partial_\mu - ig_i\mathcal{G}_{i,\mu}\right)\left(\mathbb{I}_6 + \frac{i\sqrt{2}}{f_i} \tilde{\Pi}_i - \frac{1}{f_i^2}\tilde{\Pi}_i^2 - \frac{i\sqrt{2}}{3 f_i^3} \tilde{\Pi}_i^3\right)\\
    & = g_i\mathcal{G}_{i,\mu}\\
     & + \frac{i}{f_i} \Big(i\sqrt{2}\partial_\mu \tilde{\Pi}_i - \sqrt{2}g_i[\tilde{\Pi}_i, \mathcal{G}_{i,\mu}]\Big) \\
    & + \frac{i}{f_i^2} \Big([\tilde{\Pi}_i,\partial_\mu \tilde{\Pi}_i] + ig_i[\tilde{\Pi}_i,[\tilde{\Pi}_i,\mathcal{G}_{i,\mu}]]\Big)\\
    & + \frac{i}{f_i^3} \Big(\frac{-i\sqrt{2}}{3}[\tilde{\Pi}_i,[\tilde{\Pi}_i,\partial_\mu \tilde{\Pi}_i]] + \dots \Big)\\
     & + \mathcal{O}\left(1/f^4_i\right).    
\end{align*}
with $\mathcal{G}_{i,\mu} = B_\mu, W_\mu,\mathcal{V}_\mu$, $\mathcal{A}_\mu$ with the coupling $g_i$ depending
on the factor in the product group, see also 
\cref{eq:cov_der0} and \cref{eq:cov_der1}. 
These terms lead to interaction terms in the Lagrangian containing at most four fields. 

The ones relevant for the DM investigations of this paper have the following Lorentz structures: 
\begin{equation*}
    \pi\partial_\mu\pi V^\mu \,,\quad \pi V_\mu V^\mu \,,\quad \pi\pi V_\mu V^\mu \,,\quad \pi \pi\partial_\mu\pi V^\mu \,,\quad \partial_\mu\pi\partial^\mu\pi\pi\pi\,.
\end{equation*}
The last two terms are dim-5 and dim-6 which are suppressed by corresponding powers of $f_\pi$ and therefore have little impact. 

The interactions of two pNGBS with one spin-1 resonance are characterized by the coupling
\begin{equation}
    g_{V\pi\pi} = \frac{\tilde{g} f_K^2 (r^2 - 1)}{f_\pi^2},
\end{equation}
which will get additional contributions from the electroweak gauge interactions. Last but not least, we have checked that no anomaly-induced couplings of $\ZZ$-odd pNGB to two vector bosons exist.

\bibliography{refs}
\end{document}